%% file: Ankit_Dangi_Thesis_Financial_Portfolio_Optimization_Computationally_Guided_Agents_to_Investigate_Analyse_and_Invest.tex
\documentclass[11pt, a4paper]{report}

\usepackage{natbib}
\usepackage{url}
\usepackage{graphicx}
\usepackage{amsmath}
\usepackage{pdfpages}
\usepackage{setspace}

\usepackage{authblk}
\usepackage[tmargin=1in,bmargin=1in,lmargin=1.25in,rmargin=1.25in]{geometry}

\usepackage{float}
\floatstyle{ruled}
\newfloat{algorithm}{htbp}{loa}
\floatname{algorithm}{Algorithm}

\newenvironment{ad_itemize}
{\begin{itemize}
\setlength{\itemsep}{1pt}
\setlength{\parskip}{0pt}
\setlength{\parsep}{0pt}}
{\end{itemize}}

\begin{document}





\pagenumbering{roman}
\includepdf{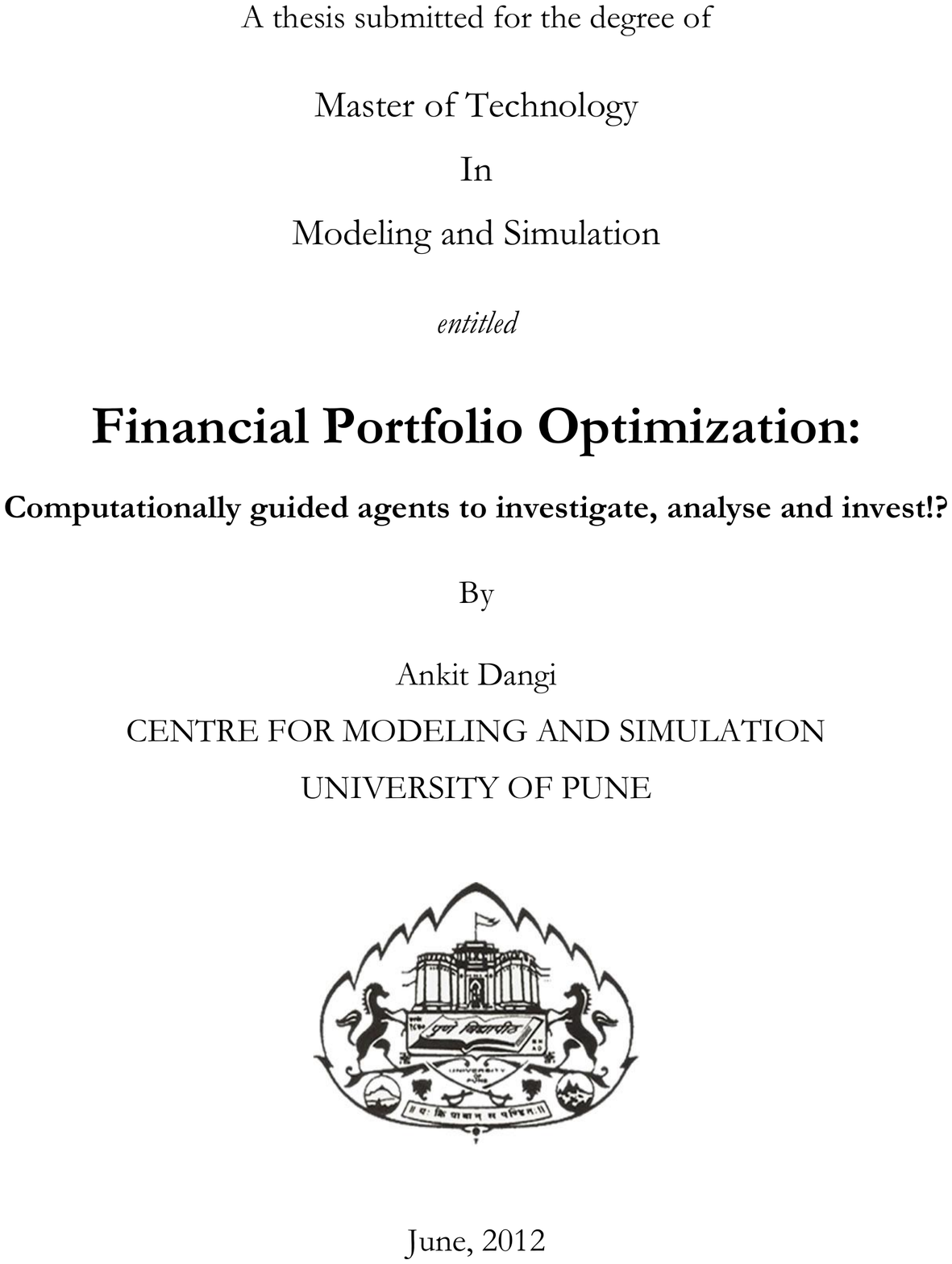}
\includepdf{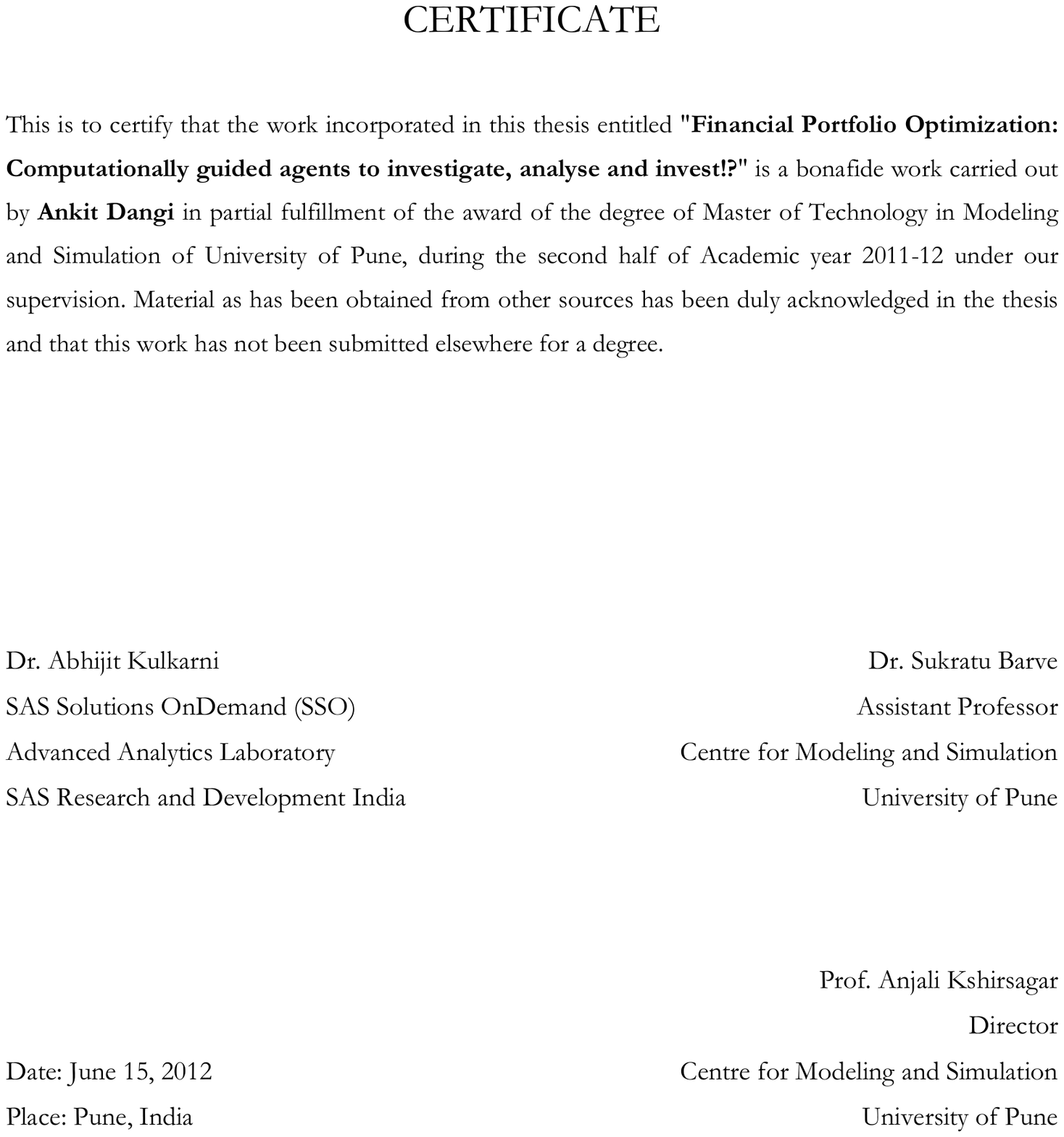}
\includepdf{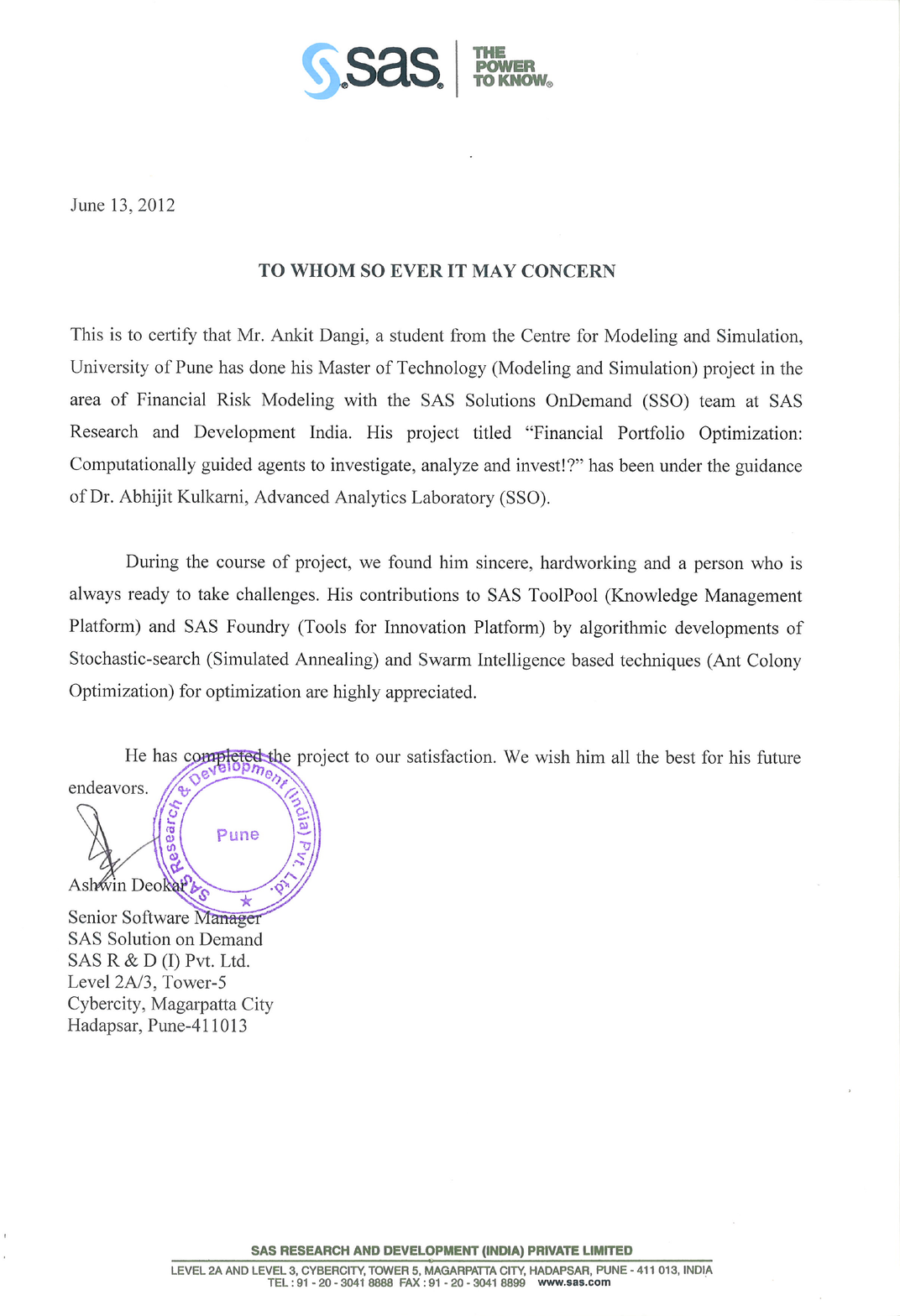}
\includepdf{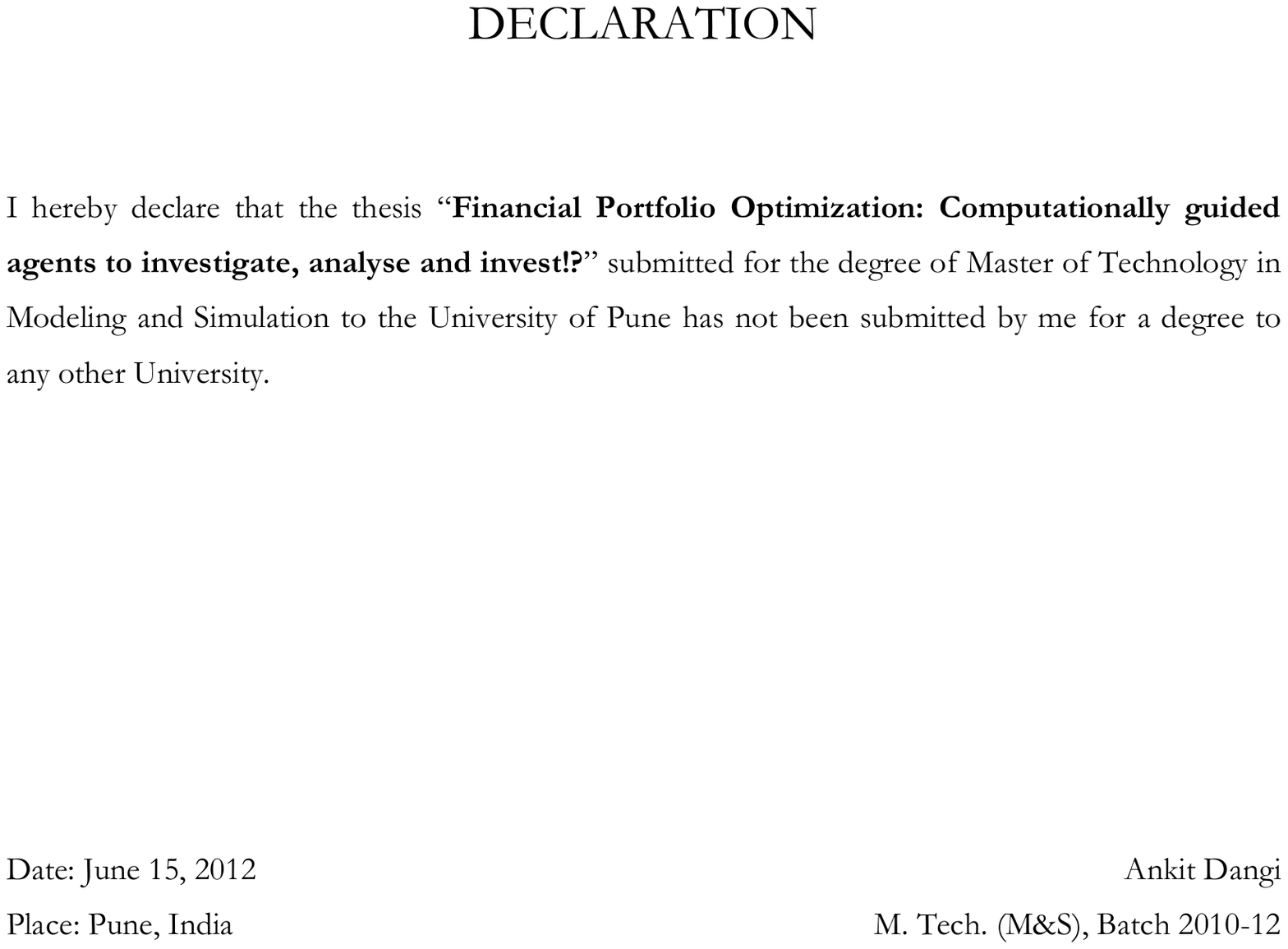}

\onehalfspacing
\include{front/acknowledgements} 
\singlespacing

\tableofcontents
\onehalfspacing

\cleardoublepage
\include{front/abstract}

\include{front/contributions}

\include{front/notations}


\listoffigures

\pagenumbering{arabic}
\singlespacing
\include{chapters/chap-1-portfolio-optimization}

\include{chapters/chap-2-problem-formulations}

\include{chapters/chap-3-computationally-guided-agents}

\include{chapters/chap-4-empirical-validation-results}

\include{chapters/chap-5-conclusions-future-work}

\appendix
\include{appendix/appendix-1}

\include{appendix/appendix-2}

\include{appendix/appendix-3}

\bibliographystyle{plainnat}
\bibliography{bibliography/biblio}

\end{document}

%% file: front/acknowledgements.tex
\chapter*{Acknowledgements}
\label{fr:acknowledgements}

Words wouldn't suffice for the deep sense of gratitude with which I would like to sincerely thank my guide Dr. Abhijit Kulkarni for his guidance throughout the course of this work. 

\paragraph{}
I would like to whole-heartedly acknowledge Prof. Uttara V. Naik-Nimbalkar and Dr. T. V. Ramanathan from Department of Statistics, University of Pune.

\paragraph{}
I am grateful to Dr. Mihir Arjunwadkar, Dr. Sukratu Barve and Prof. Anjali Kshirsagar from Centre for Modeling and Simulation, University of Pune.

\paragraph{}
I am thankful to Dr. Lokesh Nagar, Mr. Viswanath Pothinindi, Dr. Sourish Das, Mr. Prasad Paranjpe, Mr. Gaurav Singh, Dr. Swarup De, Mr. Mousum Dutta and Mr. Ashwin Deokar from SAS Solutions OnDemand, SAS Research and Development India.

\paragraph{}
I am thankful to my parents, family and friends for their support and encouragement.

%% file: front/abstract.tex
\chapter*{Abstract}
\label{fr:abstract}

Financial portfolio optimization is a widely studied problem in mathematics, statistics, financial and computational literature. It adheres to determining an optimal combination of weights that are associated with financial assets held in a portfolio. In practice, portfolio optimization faces challenges by virtue of varying mathematical formulations, parameters, business constraints and complex financial instruments. Empirical nature of data is no longer one-sided; thereby reflecting upside and downside trends with repeated yet unidentifiable cyclic behaviours potentially caused due to high frequency volatile movements in asset trades. Portfolio optimization under such circumstances is theoretically and computationally challenging. This work presents a novel mechanism to reach to an optimal solution by encoding a variety of optimal solutions in a solution bank to guide the search process with regard to the global investment objective formulation. It conceptualizes the role of individual solver agents that contribute optimal solutions to a bank of solutions, and a super-agent solver that learns from the solution bank, and, thus reflects a knowledge-based computationally guided agents approach to investigate, analyse and reach to optimal solution for informed investment decisions. 

\paragraph{}
Conceptual understanding of classes of solver agents that represent varying problem formulations and, mathematically oriented deterministic solvers along with stochastic-search driven evolutionary and swarm-intelligence based techniques for optimal weights are discussed in this work. Algorithmic implementation of the computational guidance approach from a bank of optimal solutions is presented by an enhanced neighbourhood generation mechanism in the Simulated Annealing algorithm. A framework for inclusion of heuristic knowledge and human expertise from financial literature related to investment decision making process is reflected via the introduction of controlled perturbation strategies using a decision matrix for neighbourhood generation. Empirical validation of the proposed methodology has been carried out for Bearish and Bullish market scenarios.

%% file: front/contributions.tex
\chapter*{Contributions}
\label{fr:contributions}

\section*{Academic/Computational Contributions}
\begin{itemize}

\item Computationally guided agents approach to financial portfolio optimization that incorporates varying problem formulations, variety of parameters, and complex business constraints.

\item Conceptualization of a computational model comprising of various classes of solver agents that contribute optimal solutions to a bank of solutions that guides a super-agent solver towards obtaining optimal solutions for a global investment objective.

\item Development of a novel mechanism in Simulated Annealing algorithm for neighbourhood generation using solution bank approach.

\item Conceptualization of a perturbation strategy using decision matrix for controlled disturbances to decisions variables during neighbourhood generation.

\end{itemize}

\paragraph{}
\section*{Contributions to SAS Research and Development}
\begin{itemize}
\item Portfolio Optimization Component of Asset Performance Management Solution
	\begin{itemize}
	\item[-] Development of Robust Portfolio Optimization Formulation
	\item[-] Development of Risk-based Asset Allocation Strategies 
	\end{itemize}	
\item Platform for Innovation (SASFoundry) and Knowledge Management (SAS ToolPool)
	\begin{itemize}
	\item[-] Simulated Annealing for Global Optimization (Algorithm Development)
	\item[-] Ant Colony Optimization for Continuous Domains (Algorithm Development)
	\item[-] A Suite of Benchmark Functions for Global Optimization
	\end{itemize}
\end{itemize}

%% file: front/notations.tex
\chapter*{Notations}
\label{fr:notations}

\begin{equation}
\begin{aligned}
& n/N 			&& \text{number of assets} \\
& x				&& \text{n x 1 column-vector of portfolio weights} \\
& x^T			&& \text{transpose of} \; x \; \text{; 1 x n row-vector of portfolio weights} \\
& x_i			&& i^{th} \; \text{element of} \; x \text{, denoting the weight of asset} \; i \; \text{in the portfolio} \\
& r_p			&& \text{expected return of the portfolio} \\
& \sigma_p^2 	&& \text{variance of the portfolio} \\
& r				&& \text{n x 1 vector of expected returns for} \; n \; \text{assets} \\
& s				&& \text{n x 1 vector of expected risks for} \; n \; \text{assets} \\
& r_i			&& i^{th} \; \text{element of} \; r \text{, denoting the expected return of} \; i^{th} \; \text{asset} \\
& s_i			&& i^{th} \; \text{element of} \; s \text{, denoting the expected risk of} \; i^{th} \; \text{asset} \\
& \Sigma 		&& \; n \; x \; n \; \text{covariance matrix for the returns on the assets in the portfolio} \\
& \rho_{ij} 	&& \text{correlation coefficient between returns on asset} \; i \; \text{and} \; j \; \in [-1, 1] \\
& \Gamma		&& \text{budget of uncertainty of constraints for all assets in the portfolio} \; \in [0, n] \\
& p				&& \text{uncertainty optimization parameter for budget of uncertainty} \; \Gamma \; \text{parameter} \\
& q_i			&& \text{uncertainty optimization parameter for expected risk of asset} \; i \; \text{in the portfolio} \\
\end{aligned}
\end{equation}

%% file: chapters/chap-1-portfolio-optimization.tex
\chapter{Portfolio Optimization}
\label{ch:chap-1-portfolio-optimization}

\section{Introduction, Background \& Motivation}
\label{sec:introduction}

\paragraph{}
In financial literature, a portfolio is considered as an appropriate collection of investments held by an individual or a financial institution. These investments or financial assets constitute shares of a company (often referred as equities), government bonds, fixed income securities, commodities (such as Gold, Silver, etc.), derivatives (incl. options, futures and forwards), mutual funds, and, various mathematically complex and business driven financial instruments. The individual responsible for making investment decisions using the money (or capital) that individual investors or financial institutions have placed under his/her control is referred as the Portfolio Manager. In principle, a portfolio manager holds responsibility for managing the asset and liability portfolios of a financial institution.

\paragraph{}
From a very simplistic viewpoint, consider we have a capital of One Lakh Rupees to invest in equities. Further, consider that there exists a market that has three equities: Infosys, Tata Steel and Reliance Industries. An investment perspective on the three equities raises the following fundamental questions:

\begin{ad_itemize}
\item In which of these would we invest?
\item How much would we invest in each of them?
\end{ad_itemize}

\paragraph{}
The proportion of the entire amount of investment that is divided amongst the three equities would address both the above questions. i.e. if X1 amount of money were to be invested in Infosys, X2 amount of money were to be invested in Tata Steel and X3 amount of money were to be invested in Reliance Industries such that the entire One Lakh Rupees is completely invested, then we are looking at enforcing a budget constraint such that X1 + X2 + X3 = 100\% of the entire amount of investment. Hence, if we are able to precisely determine the amount to be invested in each of these, i.e. X1, X2 and X3, then, we are determining an optimal set of weights that would correspond to equities we considered to invest. Fundamentally, determining this optimal structure of weights is considered as the Portfolio Optimization problem in Mathematical and Financial literature. In principle, the portfolio may consist of any of the complex investment options available, and would have a variety of realistic constraints.

\pagebreak
\paragraph{}
Formally, financial portfolio optimization adheres to a formal approach in making investment decisions:
\begin{ad_itemize}
\item[-] for selection of investment portfolios containing the financial instruments, 
\item[-] to allocate a specified capital over a number of available assets, 
\item[-] to meet certain pre-defined objectives, 
\item[-] to mitigate financial risks and ensure better preparedness for uncertainties, 
\item[-] to establish mathematical and computational methods on realistic constraints, 
\item[-] to manage profit and loss of the portfolio (across long and short positions), and, 
\item[-] to provide stability across inter and intraday market fluctuations etc.
\end{ad_itemize}

\paragraph{}
Banks, fund management firms, financial consulting institutions and large institutional investors, on a continual basis and repeated time-frames, are faced with the challenges of managing their funds, assets and stocks towards selecting, creating, balancing, and evaluating optimal portfolios. Markowitz Modern Portfolio Theory (MPT) has provided a fundamental breakthrough towards strengthening the mean-variance analysis framework. Ever since then, modifications, extensions and alternatives to MPT have been worked out to simplify and prioritize assumptions of the theory and to address the limitations of the framework. Parallel works to incorporate various other challenges have been introduced by numerous research papers consistently. 

\paragraph{}
Financial crisis, economic imbalances, algorithmic trading and highly volatile movements of asset prices in the recent times have raised high alarms on the management of financial risks. Inclusion of risk measures towards balancing optimal portfolios has become very crucial and equally critical. In recent times, varied mathematical models have emerged leading towards practical risk-based asset allocation strategies. Undoubtedly, incorporation of risk factors brings additional realistic constraints with an increased number of parameters towards formulation of mathematically complex and higher order objective functions that challenge the role of deterministic mathematical optimization routines. 

\paragraph{}
In today's world, technological advances to computation speed, power and volume needs minimal introduction. Powerful computing infrastructure that account for high-performance computing, grid-based analytics and in-memory analysis of large volumes of data lay strong basis for the role of computational modeling in the world of finance. And, larger possibilities of uncertain events, growing number of markets, mathematically modeled complex financial instruments etc. are some of the few challenges that computational models and algorithms tend to address. 

\paragraph{}
The diffusion of low-cost high-performance computing capabilities have allowed for the broad use of numerical methods. The importance of finding closed-form solutions and the consequent search for simpler models, and when required, complex models provide stronger emphasis for computationally-intensive methods such as Monte Carlo simulations, numerical approximations to differential equations (ordinary and stochastic), population based approaches, heuristic branching of uncertain possibilities in a search space etc. provide just a gist of a variety of possibilities that computational sciences tend to offer. Addressing the availability of such high-valued computing techniques, and to overcome challenges faced by deterministic optimization methods, this work does focus towards having a look at stochastic-search based optimization routines towards optimal asset allocation strategies.

\paragraph{}
Interestingly, deeper understanding of the mechanisms of financial markets seems to suggest that they tend to behave similar to the systems, processes and phenomena that occur in the natural world around us. In computational sciences, treatments to complex problems via robust algorithms have repeatedly taken metaphorical inspirations from nature-inspired methods of reaching optimal solutions, just like the way, bio-organisms explore vast number of possibilities to reach to an optimal solution. Physical processes that occur in nature assist in providing solutions via repeated trials and facilitate in reducing estimation errors for the decision variables of objective functions. And, evolutionary approaches in nature strengthen the stability of the solution across trials from one generation to another.

\paragraph{}
In light of the above introduction and background, the next section inclines towards exploring further insights from literature and then takes a re-look at the problem of financial portfolio optimization.

\section{Literature Survey}
\label{sec:literature}

\paragraph{}
Formal approach towards making investment decisions for obtaining an optimal portfolio with a specific objective requires a mathematical formulation for the problem. Validation of the models from a mathematical perspective is undoubtedly challenging and requires rigorous understanding. However, with increasing computational capabilities and the growing volumes of data, computational models and algorithms are often developed to validate the mathematical models with empirically available data. Standing on the shoulders of the giants, this section intends to explore their work by providing a literature survey on the theoretical developments and computational approaches to the problem. 

\paragraph{}
Fundamental breakthrough in the problem of asset allocation and portfolio optimization is often dated to the Markowitz's Modern Portfolio Theory \citep{Markowitz1952}. It considers rational investors and models the problem keeping the mean-variance analysis framework in perspective, where, the variance of the portfolio is minimized with a fixed value for the expected return on the entire portfolio. The framework also assumes a market without any taxes or transaction costs, and where short selling is disallowed but assets are infinitely divisible and can be traded with any non-negative fractions. 

\paragraph{}
Tobin James's work \citep{Tobin1967} presents the inclusion of risk-free assets in the traditional Markowitz formulation by the development of the Separation theorem which states that in the presence of a risk-free asset, the optimal risky portfolio can be obtained without any knowledge of the investor's preferences, whereas, Sharpe's Capital Asset Pricing Model (CAPM) \citep{Sharpe1964} takes into account the asset's sensitivity to non-diversifiable risk while it is being added to an already existing well-diversified portfolio. It considers the importance of the covariance structure of the returns, the variance of the portfolio and the market premium (i.e. the difference between the expected return on the asset from the market and the risk-free rate of return on the asset). The model assumes that the investors are rational and risk-averse, are broadly diversified across a range of investments, and that they cannot influence the prices of the assets. Assumptions regarding trade or transaction costs, short-selling and trades with non-negative fractions do apply from the traditional Markowitz's framework.

\paragraph{}
More recently, Ross's Arbitrage Pricing Theory (APT) \citep{Ross1976} models the expected return of a financial asset as a linear function of various macro-economic factors (where, sensitivity to changes in each factor is represented by the factor-specific beta coefficients of the regression model) and Miller's Cost of Capital \citep{Miller1958} or the Capital Structure Irrelevance Principle forms the foundation for looking at the capital structure. Mathematically, referred as Modigliani-Miller Theorem \citep{wiki:ModiMiller2012}, it states that under a certain market the price process is unaffected by the financing structure of the firm. The theorem assumes the absence of trade and transaction costs, taxes, and assumes presence of information asymmetry and an efficient market. 

\paragraph{}
Mathematical/Statistical perspectives to the problem of portfolio optimization range across various other schools of thought as well. Over the years, tremendous advances in literature \citep{IIF2009}, \citep{Balbas2007}, \citep{Distortion2010} and \citep{Ortobelli2005} have been in place with regard to exploring return and risk measures for quantifying the parameters of the portfolio optimization problem including constant and time-varying higher moments on the returns. In order to characterize the uncertainties in parameter estimates, Michaud's resampled efficiency technique \citep{Michaud2008} lays emphasis on continuous re-sampling from empirical data. Approaches from mathematical physics, include the use of asset selection filters based on Random Matrix Theory as mentioned in \citep{RMT2004}, \citep{Daly2008} and \citep{Crane2007} are being applied to covariance structures of the returns on the assets to provide stability to the problem and provides mechanisms for assessment of risks. Bootstrapping techniques are often considered in such approaches.

\paragraph{}
In studies of financial calculus, considering the equity markets in perspective, Fernholz’s Stochastic Portfolio Theory \citep{Fernholz2002} and \citep{Fernholz2008} discusses a descriptive theory that provides a framework for analysing portfolio behaviour and equity market structure that has both theoretical and practical applications. It provides insights into questions of market behaviour and arbitrage principle. It is used to construct portfolios with controlled behaviour under certain conditions and provides methods for managing the performance of the portfolio. 

\paragraph{}
From a behavioural science and economic perspective, Von Neumann-Morgenstern's Expected Utility Hypothesis \citep{Wiki:VNM2012} plays a crucial role in looking at the investor's betting preferences with regard to uncertain price movements being represented by a function of the payouts, the probabilities of occurrence, risk aversion, and the different utilities of the same payout to other investors with different assets or personal preferences. It tends to provide an alternative perspective at considering the expected value approach referred in statistics by focusing on four fundamental axioms that would characterize a rational decision maker. These axioms include completeness, transitivity, independence and continuity. Various works including \citep{Neumann1953}, \citep{Marschinski2007} and \cite{Markowitz2010} discusses the inclusion of such a hypothesis while considering the portfolio optimization problem. 

\paragraph{}
Over the recent years, with regard to various social media information being generated on a continual basis, Bayesian perspective to the problem seems to be gaining popularity. Black-Litterman's model \citep{BlackLitterman1992} considers including the views of the investors as prior information into the estimates of the expected asset returns and the covariance structure. It provides a mechanism to model user-specified confidence levels that are based on the strong/weak opinions of the investors and financial experts, either in totality or with partial knowledge, and tends to span across arbitrary and/or overlapping sets of assets. Related works that discuss the Bayesian approach to the portfolio optimization problem include \citep{Christodoulakis2002}, \citep{Hoffman2011}, \citep{Litterman1999}, \citep{Walters2009} and \citep{Zhou2009}.

\paragraph{}
The above portfolio optimization theories and various other theoretical advances have been proposed with the aid of mathematical and statistical modeling methods. For varied surveys on advances in literature, one is referred to \citep{Nishimura1990}, \citep{Bolshakova2009}, \citep{Moore1972} and \citep{Figueroa2005}.

\paragraph{}
Considering the wide range of advances in the field of computational sciences in terms of multiple paradigms for computations, availability of high performance computing infrastructure and methodical simulation strategies, analysis and empirical validation of mathematical and statistical models is inclined towards exploring these enhanced computational capabilities. Further literature presented in this section, intends to broadly highlight computational developments without presenting mathematical variants of the portfolio optimization problem that have been considered in them. The spectrum of such advancements range from conventional numerical methods of analytic and non-analytical approximations, to stochastic programming via simulations exploring multi-period scenario generations, stochastic optimization and stochastic-search based optimization using local search procedures, evolutionary computations and swarm intelligence using population based algorithmic developments, and, machine learning based computationally guided mechanisms.

\paragraph{}
Formulation of Markowitz's portfolio optimization problem is viewed upon as a quadratic optimization problem. \citep{Boyd2009} and \citep{Nocedal1999} provides a comprehensive literature to convex and numerical optimization methods to solve such a formulation. A commonly known pitfall for numerical approximations via computational systems is viewed upon by finite precision arithmetic including both the fixed-point and the numerical errors in floating-point arithmetic. \citep{Antia1991} and \citep{Press2007} provides a detailed overview of scientific computing methods and \citep{Fernando2000} considers a practical cum numerical perspective on the portfolio optimization problem.

\paragraph{}
In the literature corresponding to stochastic programming and simulations, \citep{Yu2003} provides a survey of stochastic programming models. \citep{Parpas2006} explores a global optimization approach to scenario generation and portfolio optimization looking at them as individual problems. \citep{Geyer2009} proposes a stochastic programming approach for multi-period portfolio optimization. \citep{Deniz2009} presents a multi-period scenario generation approach to support portfolio optimization and \citep{Guastaroba2010} discusses scenario generation, mathematical models and algorithms for the portfolio optimization problem. \citep{Troha2011} proposes the problem as stochastic programming with polynomial decision rules by presenting coherency and time-consistency as risk measures for the problem. And, \citep{Greyserman2006} explores portfolio selection using hierarchical Bayesian analysis and Markov Chain Monte Carlo (MCMC) methods.

\paragraph{}
Stochastic optimization methods refer to optimization models where random variables appear in the problem formulation. These methods tend to appear as solution approaches to portfolio optimization problem modeled with reference to the Stochastic Portfolio Theory as mentioned above. Stochastic-search based optimization using local search procedures refers to introducing randomness in to the search process to accelerate progress towards leading to the optimal local solution. Challenges in such a methodology are with regard to handling constraints. \citep{Crama2003} provides a simulated annealing based approach to complex portfolio selection problems by a systematic inclusion of constraints. \citep{Ardia2010} and \citep{Krink2011} provide differential evolution based stochastic-search heuristic methods for multi-objective portfolio optimization problem with realistic constraints. 

\paragraph{}
Computational methods inspired by biological mechanisms of evolution including reproduction, mutation, recombination and selection are referred in evolutionary computation. They use iterative progression amongst a population of individuals, select them for guided random search and process them in parallel. Evolutionary algorithms are population-based meta-heuristic optimization methods, and \citep{Hochreiter2008} performs evolutionary stochastic portfolio optimization using genetic algorithm as the meta-heuristic. \citep{Branke2009} discusses the portfolio optimization with an envelope-based multi-objective evolutionary algorithm with a variety of non-convex constraints. \citep{Roudier2007} designs a multi-factor objective function reflecting investment preferences and solves the portfolio optimization problem using genetic algorithm. \citep{Chan2002} applies genetic algorithms in multi-stage portfolio optimization system, and \citep{Lin2005} solves the problem with the same method but considers transaction costs and minimum transaction lot constraints.

\paragraph{}
Computational algorithms that believe in collective behaviour of systems which are decentralized and self-organized are referred as swarm intelligence based algorithms. \citep{Doerner2001} considers the use of meta-heuristics towards performing an ant colony optimization in the multi-objective portfolio selection problem and \citep{Thong2007} presents constrained Markowitz portfolio selection using ant colony optimization. \citep{Mishra2009} considers multi-objective particle swarm optimization approach to the portfolio optimization problem, and \citep{Chen2006} considers particle swarm optimization for constrained portfolio selection problems. \citep{Karaboga2009} and \citep{Karaboga20091} provide a comprehensive survey of algorithms that simulate bee swarm intelligence algorithms and its applications. \citep{Vassiliadis2008} applies the artificial bee colony optimization algorithm for the constrained portfolio optimization problem. An experimental study on a hybrid nature-inspired algorithm considering the ant colony optimization and the firefly algorithm is proposed in \citep{Giannakouris2010} for financial portfolio optimization. Various other swarm intelligence based techniques include cuckoo search, stochastic diffusion search, artificial immune systems, charged system search, and many more.

\paragraph{}
Machine learning based methods that refer to statistical learning with data are widely applicable in computational finance. \citep{Still2010} provides a mechanism for regularizing the portfolio optimization that uses the L2 norm of the weight vector as a regularizer to act upon as a pressure towards portfolio diversification and resembles the same to support vector regression. \citep{Diagne2002} discusses financial risk management and portfolio optimization using artificial neural networks and extreme value theory. \citep{Zimmermann2001} performs active portfolio management based on error correction neural networks where the problem is modeled as a feed-forward neural network and the underlying expected return forecasts are based on error correction neural networks. \citep{Steiner1997} considers portfolio optimization with a neural network implementation of the coherent market hypothesis. And, \citep{Zhang2008} presents the portfolio optimization with predictive distribution using dynamic graphical models including vector autoregressive, Kalman filters and proposes Sparse-Gaussian Kalman filters to regularize the inverse covariance of the state and observation noise to reduce the variance of model estimation and the risk associated with the resulting portfolios.

\paragraph{}
The theoretical advances and computational techniques that appear in literature on financial portfolio optimization reflect the on-going and progressive work in this field. The plethora of information available cites the relevance of the asset allocation and portfolio optimization problem in financial studies. This work may not be able to do complete justice in exploring the literature to as exhaustive and comprehensive it could be, but it believes to have covered the broader spectrum of available literature. Considering the challenging nature of market dynamics, and growing complexities, this work focuses to contribute to the pool of knowledge by looking at the problem from a computational perspective.

\section{Problem Statement}
\label{sec:problem-statement}

\paragraph{}
Mathematical modeling of the portfolio optimization problem attempts to consider various parameters of the problem addressing towards return measures and more recently, even risk measures. Considering the significance of the challenges faced in the management of portfolio, theoreticians and practitioners constantly revive and revise various methods of measuring returns, risks and performance of portfolio. Based on the literature survey of the problem as discussed in the previous section, and considering the real-world complexities of working under financial limitations and policy considerations, the models of portfolio optimization are worked out with a number of constraints and with varying objectives.

\paragraph{}
In practice, portfolio managers of financial institutions are constantly faced with the challenges of adapting the theoretical models formulated earlier to the dynamics of the current scenarios. Each portfolio manager has a different outlook at the portfolio that he/she is responsible to manage and with the nature of empirical data; the need to relax a few assumptions of those models has gained greater importance. Some portfolio managers may allocate their entire wealth to be invested equally across all assets, whereas, some might strictly adhere to the principles formulated by Markowitz and work towards minimizing the variance of the portfolio.

\paragraph{}
A portfolio manager working with a few number of assets might believe in the diversification of the wealth across all assets by maximizing the diversification ratio of the portfolio whereas, another portfolio manager who believes in reasoning and decision making under uncertainty might tend to consider the uncertainties involved in estimating the parameters of the optimization problem. Some manager might even have a re-look at the entire formulation by considering risk-allocations across the portfolio as compared to allocations in assets thereby ensuring that the contributions to risk by all assets are neutralized. Further, each portfolio manager may also be interested in evaluating his/her portfolio's performance on a continual basis and may tend to relate the portfolio performance by associating its comparison with a relevant market driven benchmark.

\paragraph{}
Studying the plethora of diverse mathematical formulations in theory that are driven by a breadth of objective functions, a variety of realistic constraints and a varied set of parameters as described in the previous section, and considering the computational approaches that have been applied to the problem, along with strategical incorporation of challenges faced by portfolio managers in practice, the financial institution and/or individual investor is faced with the fundamental question of reaching to an optimal weight structure for the investments that would align towards the investor's global objective.

\paragraph{}
This work intends to focus on the problem by combining optimal solutions obtained by a variety of sources including formulations, solvers, strategies and perspectives to further align the investor's wealth allocation towards an optimal structure that would potentially facilitate further guidance towards informed decision making in the financial portfolio optimization circumstances and challenges. The forthcoming Chapter ~\ref{ch:chap-2-problem-formulations} explains a few mathematical formulations of the problem and Chapter ~\ref{ch:chap-3-computationally-guided-agents} discusses the proposed solution approach using computational guidance from a set of optimal solutions. Chapter ~\ref{ch:chap-4-empirical-validation-results} presents empirical validation of the proposed work via computational experiments and results are further discussed. Towards the end, in Chapter ~\ref{ch:chap-5-conclusions-future-work}, findings of this work are concluded, and it discusses future course of work that may be taken up to strengthen the proposed solution.

%% file: chapters/chap-2-problem-formulations.tex
\chapter{Problem Formulations}
\label{ch:chap-2-problem-formulations}

\section{Markowitz Modern Portfolio Theory (MMPT)}
\label{sec:markowitz}

\paragraph{}
The fundamental breakthrough towards managing financial investments was provided by Harry Markowitz by constructing a portfolio optimization formulation with the mean-variance analysis framework. In statistical terms, investments are described in terms of their expected long-term return rate and their expected short-term volatility. This formulation supports the concept of diversification of entire investment across all assets. The theory focusses on minimization of the variance of the portfolio for a specified expected portfolio return under the assumptions that are briefly mentioned in Section ~\ref{sec:literature}. In general, the mean-variance trade-off dictates that with increased levels of risks that an investor wishes to take, the investor is then expected to obtain increased returns. This associates a linear relationship between the risk levels and the expected returns.

\paragraph{Markowitz Formulation}
\begin{equation}
\begin{aligned}
& \min & \sigma_p^2 = \sum^n_{i=1} \sum^n_{j=1} x_i x_j \, s_i s_j \, \rho_{ij} 
&&& \text{objective function} \\
& \text{subject to} & \sum^n_{i=1} r_i x_i = r_p &&& \text{return constraint} \\
& & \sum^n_{i=1} x_i = 1 &&& \text{budget constraint} \\
& & 0 \leq x_i \leq 1, \; \forall i &&& \text{long-only constraint} \\
& \text{where, } & \rho_{ij} \in [-1, 1], \; \forall i, j &&& \text{correlation coefficient}
\end{aligned}
\end{equation}

\paragraph{Markowitz Formulation (Linear Algebra Form)}
\begin{equation}
\begin{aligned}
& \min & \sigma_p^2 = x^T \Sigma x &&& \text{objective function} \\
& \text{subject to} & \sum^n_{i=1} r_i x_i = r_p &&& \text{return constraint} \\
& & \sum^n_{i=1} x_i = 1 &&& \text{budget constraint} \\
& & 0 \leq x_i \leq 1, \; \forall i &&& \text{long-only constraint} \\
\end{aligned}
\end{equation}

\paragraph{}
In order to avoid the slow or infeasible convergence of the above optimization formulation, this work relaxes the return constraint from a hard constraint to a soft constraint.

\paragraph{Markowitz Formulation (Soft Return Constraints)}
\begin{equation}
\begin{aligned}
& \min & \sigma_p^2 = x^T \Sigma x &&& \text{objective function} \\
& \text{subject to} & \sum^n_{i=1} r_i x_i \ge r_p &&& \text{return constraint} \\
& & \sum^n_{i=1} x_i = 1 &&& \text{budget constraint} \\
& & 0 \leq x_i \leq 1, \; \forall i &&& \text{long-only constraint} \\
\end{aligned}
\end{equation}

\paragraph{}
Where, the asset returns considered for the above formulation are the absolute return measures and absolute risk measures. The formulation considers arithmetic returns for the assets as the return measure and the standard deviation as the risk measure. 

\paragraph{}
Additionally, one could consider formulating the problem with other classes of parameters of estimating return measures that would address critical aspects of the financial markets which are significant to large financial institutional investors and public/private banks. These may include the Average Profit and Loss i.e. Average PnL (as an absolute value, or as a \%age, or per each industrial sector, or as per a pre-specified time interval), and Compounded Annual Growth Rate (CAGR). Other critically relevant measures may include the relative return measures (incl. upward-downward movement capture ratio, or the upward-downward movement number/percentage ratio), or the absolute risk-adjusted return measures (incl. Sharpe ratio, Calmar ratio, Sterling ratio or Sortino ratio), or the relative risk-adjusted return measures (incl. Annualized alpha, Jenson alpha, Treynor ratio, or Information ratio).

\paragraph{}
In practice, economic policies and financial trades impose a variety of constraints on the traditional formulation. Relevance of the budget constraint and the return constraint seem natural to the problem. Further imposition of constraints such as holding constraints, risk-factor (risk-fraction) constraints, trading (transaction size) constraints, cardinality constraints, round lot constraints, volatility constraints, closing position constraints, turnover (purchase/sale) constraints, tracking error constraints etc. tend to various multiple formulations of the portfolio optimization problem. For a detailed overview of the potential parameters and feasible constraints usually taken into consideration, refer Appendix ~\ref{ch:appendix-1} and ~\ref{ch:appendix-2}.

\section{Robust Formulation addressing Parameter Uncertainty}
\label{sec:robust}

\paragraph{}
The presence of a statistical measure for estimating returns, risk and the covariance matrix of asset returns that are parameters to the mathematical model are subject to uncertainty with respect to the expected measure as compared to the actual one. Practical reasons for such uncertainties include minor fluctuations to the financial data, lack of availability of accurate information for very small time points, and the existence of uncertain and often unknown factors. In practice, sensitivity analysis does address these issues to some extent but tends to be well-suited only after an optimal solution is obtained. In order to handle such critical relevance of parameter uncertainty during the computation of optimality it is significant from a mathematical perspective, as well, because imperfect information to the model does tend to threaten the relevance of the solution.

\paragraph{}
To address parameter uncertainty issues, methods in computational statistics literature suggest the formulation of the problem as stochastic processes to inherently handle the stochastic behaviour of the asset returns, or, to deal with the stochastic uncertainty (due to randomness in the model) over multiple stages of the solution space search. The former is regarded as stochastic programming whereas the latter is referred as dynamic programming. However, obtaining the probability distributions of the uncertainties in the model poses tactical difficulties towards the stochastic programming approach, and, the lookout for an optimal solution in a very large search space due to scenario generation methods of dynamic programming highlights the practical difficulties of computational expensiveness, greater costs, and smaller revenues for optimality.

\paragraph{}
An alternative method, acknowledged to \citep{Soyster1973}, considers modeling of every uncertain parameter as taken equivalent to its worst-case value within a set of feasible parameters. This method is inspired from the robust control engineering literature, and is hence, referred as Robust Optimization. Such a methodology does address the problem but tends to be a bit too conservative. And, to work around the over-conservatism, the uncertain parameters are restricted to belong to an uncertainty set. Research in the 90's by \citep{BenTal1998}, \citep{BenTal1999}, \citep{BenTal2000}, \citep{ElGhaoui1997} and \citep{ElGhaoui1998}, suggest the use of ellipsoidal uncertainty sets, whereas, more recently, \citep{Bertsimas2003}, \citep{Bertsimas2004} and \citep{Bertsimas20041} have proposed the robust optimization approach based on polyhedral uncertainty sets to preserve the class of problems under analysis i.e. a linear programming problem stays as a linear programming problem as compared to the ellipsoidal set where the linear programming problem transforms to a second-order cone problem. 

\paragraph{Robust Optimization}
\begin{equation}
\begin{aligned}
& \max & \sum^n_{i=1}r_i x_i - \Gamma_i p_i - \sum^n_{j=1} q_{ij} &&& \text{objective function} \\
& \text{subject to} & \sum^n_{i=1} x_i = 1 &&& \text{budget constraint} \\
& & p_i + q_{ij} \geq s_i x_i, \; \forall i &&& \text{uncertainty constraint} \\
& & 0 \leq x_i \leq 1, \; \forall i &&& \text{long-only constraint} \\
& & p_i, q_i \geq 0, \; \forall i &&& \text{non-negativity constraint} \\
\end{aligned}
\end{equation}

\paragraph{Robust Optimization (Simplified)}
\begin{equation}
\begin{aligned}
& \max & \sum^n_{i=1}r_i x_i - \Gamma p - \sum^n_{i=1} q_{i} &&& \text{objective function} \\
& \text{subject to} & \sum^n_{i=1} x_i = 1 &&& \text{budget constraint} \\
& & p + q_i \geq s_i x_i, \; \forall i &&& \text{uncertainty constraint} \\
& & 0 \leq x_i \leq 1, \; \forall i &&& \text{long-only constraint} \\
& & p, q_i \geq 0, \; \forall i &&& \text{non-negativity constraint} \\
\end{aligned}
\end{equation}

\paragraph{}
It can be seen that the above robust formulation of portfolio optimization resembles the Markowitz's formulation except that the portfolio risk (standard deviation) is being penalized instead of the variance. However, one may note that this robust optimization technique towards performing mathematical optimization under uncertainty considers an uncertainty model that is deterministic. Theoretically, for different types of uncertainty sets, in order to obtain the uncertain returns, one can define different risk measures on the expected portfolio return. In this way, there seems to be a natural link between the robust optimization methodology and the mean-risk portfolio optimization.

\section{Risk-based Asset Allocation Strategies}
\label{sec:rbaa}

\paragraph{}
Asset allocation strategies in theory are broadly classified as: Strategic, Tactical and Dynamic. The strategic asset allocation methods address to create an asset mix that provides an optimal balance between expected returns and risk, and are often, preferred for the long run whereas, tactical asset allocation strategy takes a more active approach by positioning a portfolio into those assets, sectors or individual stocks that portray most potential for gains and is preferred for short-term profits. Tactical asset allocation is more suited towards high-performance trading scenarios. And, the dynamic asset allocation focusses on creating a constant mix of assets as markets rise/fall and as the economy strengthens/weakens. It believes in taking short positions in the declining assets, and taking long positions in the increasing assets. Other asset allocation strategies include insured, integrated etc. 

\paragraph{}
In recent years, the global financial crisis has led to numerous formulations with a growing amount of literature on effective management and optimization of financial portfolios via Strategic asset allocation. \citep{Lee2011} suggests that many studies have been attributed to better performance of strategic risk-based asset allocation techniques as compared to conventional methods of superior diversification. However, as discussed in Section ~\ref{sec:problem-statement}, given the absence of a well-defined objective function for investment management and portfolio optimization based on the strategic approaches as well as the metrics used to evaluate the portfolio performance, \citep{Lee2011} approaches the problem into the traditional context of mean-variance efficiency in an attempt to understand their theoretical underpinnings. Some of these portfolio strategies of strategic risk-based asset allocation are briefed below.

\paragraph{Equally-Weighted (EW) Portfolio aka. 1/N Portfolio:}
This strategy believes in diversification by allocating equal weights to all assets in the composition of the portfolio. If the number of assets in the portfolio is large, then, smaller weights are allocated to the assets. Hence, it is the least concentrated in terms of weights. Such a portfolio management strategy ignores the characteristics of the assets, does not apply any constraint other than the budget constraint, and lacks the presence of an objective function.

\begin{equation}
\begin{aligned}
& && x_i = 1/N, \; \forall i &&& \text{equal weights} \\
\end{aligned}
\end{equation}

\paragraph{Global Minimum Variance (GMV) Portfolio:}
This strategy believes in minimizing the variance of the portfolio without the presence of a return constraint. It is constructed by only considering the covariance matrix of the returns on the assets and the imposition of the budget constraint. It tends to allocate higher weights to the low-volatility assets which in-turn tend to be more sensitive to the parameter estimates for both the variances and the covariance matrix of returns. By minimizing the portfolio variance without considering expected returns, the marginal contributions to risks of all assets in the portfolio are identical and rather, equal to the volatility of the portfolio and well-concentrated. 

\begin{equation}
\begin{aligned}
& \min & \sigma_p^2 = \sum^n_{i=1} \sum^n_{j=1} x_i x_j \, s_i s_j \, \rho_{ij} 
&&& \text{objective function} \\
& & \sum^n_{i=1} x_i = 1 &&& \text{budget constraint} \\
& & 0 \leq x_i \leq 1, \; \forall i &&& \text{long-only constraint} \\
& \text{where, } & \rho_{ij} \in [-1, 1], \; \forall i, j &&& \text{correlation coefficient}
\end{aligned}
\end{equation}

\paragraph{}
\paragraph{}
or, in linear algebra form, it can be written as:

\begin{equation}
\begin{aligned}
& \min & \sigma_p^2 = x^T \Sigma x &&& \text{objective function} \\
& & \sum^n_{i=1} x_i = 1 &&& \text{budget constraint} \\
& & 0 \leq x_i \leq 1, \; \forall i &&& \text{long-only constraint} \\
\end{aligned}
\end{equation}

\paragraph{Most Diversified Portfolio (MDP):}
This strategy believes in maximizing the entire diversification of wealth across all assets in the portfolio. In principle, maximum diversification can be achieved by defining a Diversification Ratio that would maximize the distance between two volatility measures of the same portfolio or by minimizing the variance of the portfolio with a risk-weighted constraint that would allow for budgeting of the risks involved in maximizing the diversification in the portfolio. The diversification ratio approach is often referred as the Maximum Sharpe Ratio (MSR) approach, because if the Sharpe ratio is the same for all assets then maximizing the diversification ratio is equivalent to maximizing the Sharpe ratio. And, even though the ratio is maximized, the portfolio is relatively concentrated with regard to the weights and risk contributions of the assets. 

\begin{equation}
\begin{aligned}
& \max & DR = \frac{\sum^n_{i=1} s_i x_i}{\sqrt{\sum^n_{i=1} \sum^n_{j=1} x_i x_j \, s_i s_j \, \rho_{ij}}} 
&&& \text{objective function} \\
& & \sum^n_{i=1} x_i = 1 &&& \text{budget constraint} \\
& & 0 \leq x_i \leq 1, \; \forall i &&& \text{long-only constraint} \\
& \text{where, } & \rho_{ij} \in [-1, 1], \; \forall i, j &&& \text{correlation coefficient} 
\end{aligned}
\end{equation}

\paragraph{}
\paragraph{}
or, in linear algebra form, it can be written as:

\begin{equation}
\begin{aligned}
& \max & DR = \frac{x^T \sigma}{\sqrt{x^T \Sigma x}} &&& \text{objective function} \\
& & \sum^n_{i=1} x_i = 1 &&& \text{budget constraint} \\
& & 0 \leq x_i \leq 1, \; \forall i &&& \text{long-only constraint} \\
\end{aligned}
\end{equation}

\paragraph{}
The risk-weighted approach is similar to the diversification ratio approach, where risk-weights are modeled as constraints and the budget constraint is relaxed. For optimality, it allows for weights to be allocated more than the entire wealth, and to meet the financial budget constraints, it further normalizes the weights. It has been observed in practice that such a risk-based indexation promises diversification and is independent of the market capitalization.

\begin{equation}
\begin{aligned}
& \min & \sigma_p^2 = \sum^n_{i=1} \sum^n_{j=1} x_i x_j \, s_i s_j \, \rho_{ij} 
&&& \text{objective function} \\
& & \sum^n_{i=1} s_i x_i = 1 &&& \text{risk-weight constraint} \\
& & x_i \geq 0, \; \forall i &&& \text{long-only constraint} \\
& \text{where, } & \rho_{ij} \in [-1, 1], \; \forall i, j &&& \text{correlation coefficient}
\end{aligned}
\end{equation}

\paragraph{}
\paragraph{}
or, in linear algebra form, it can be written as:

\begin{equation}
\begin{aligned}
& \min & \sigma_p^2 = x^T \Sigma x &&& \text{objective function} \\
& & \sum^n_{i=1} s_i x_i = 1 &&& \text{risk-weight constraint} \\
& & x_i \geq 0, \; \forall i &&& \text{long-only constraint} \\
\end{aligned}
\end{equation}

\paragraph{Equally-weighted Risk Contribution (ERC) Portfolio:}
This strategy of asset allocation believes in minimizing the variance of rescaled risk contributions. The risk contribution of an asset is the share of the total portfolio risk that is attributable to the constituting asset. Formally, it is defined as the product of the weight associated with the asset and its marginal contribution to risk which is beta times the volatility of the underlying asset. Risk allocations and risk budgeting are commonly known terms in practice that follow such principles of analysing the portfolio in terms of the risk contributions. And, when the risk contributions of all assets are equalized, this strategy is also referred as the Risk Parity approach. In principle, it constructs a highly diversified portfolio which is least sensitive to the covariance matrix of asset returns, and the percentage contribution to risks of all assets is equal. 

\paragraph{}
\citep{Maillard2008} discusses the theoretical properties of the ERC portfolio construction strategy and mathematically, discusses two formulations of constructing the optimization problem: one, with the relative risk of the asset compared to the rest of the portfolio, and the other, as a portfolio variance minimization problem subject to a non-linear inequality constraint that would lead to sufficient diversification of weights. This work considers the first formulation. \citep{Levell2010} and \citep{Meketa2010} present the relevance of risk allocation in recent times of financial crisis.

\begin{equation}
\begin{aligned}
& \min & \sum^n_{i=1} \sum^n_{j=1} (x_i (\Sigma x)_i - x_j (\Sigma x)_j)^2
&&& \text{objective function} \\
& & \sum^n_{i=1} x_i = 1 &&& \text{budget constraint} \\
& & 0 \leq x_i \leq 1, \; \forall i &&& \text{long-only constraint} \\
& \text{where, } & \rho_{ij} \in [-1, 1], \; \forall i, j &&& \text{correlation coefficient} \\
& \text{and, } & \Sigma x = \sum^n_{i=1} \sum^n_{j=1} \sigma_{ij} x_j &&& \text{rescaled risk contributions}
\end{aligned}
\end{equation}

\paragraph{}
Looking upon the practical relevance of the risk-based asset allocation strategies, theoretical interests in these methods have been on the rise. Some approaches alike \citep{Lindberg2009} even consider modeling the returns on the assets as unobservable, independent Brownian motions with a similar drift across all assets. These Brownian motions seem to collectively represent the return characteristics incorporating the volatilities and the correlations across assets. Furthermore, equal weights are allocated to the Brownian motions. In principle, theoretically driven complex stochastic processes may be considered as representatives for the return and risk measures of the assets. To drive a little more towards reality, perhaps one may even consider identifying the exact distribution of the asset returns, and based upon which, a representative stochastic process may be simulated for each asset, and then, various tactical and strategic asset allocation mechanisms may further be considered for optimal allocation of weights.

\paragraph{}
Considering the brief exploration of various theoretical and practical problem formulations as presented above, and keeping in perspective the problem statement as described in Section ~\ref{sec:problem-statement}, the next section focusses towards addressing the financial portfolio optimization problem by combining various optimal solutions and further driving a computationally guided approach towards an acceptable solution.

%% file: chapters/chap-3-computationally-guided-agents.tex
\chapter{Computationally Guided Agents Approach}
\label{ch:chap-3-computationally-guided-agents}

\section{Conceptual Representation}
\label{sec:concept}

\paragraph{}
Previous chapters have presented the challenges faced in the area of financial portfolio optimization. Theoreticians and practitioners have constantly worked around this problem by proposing various formulations using business and real-world constraints that seem applicable to the portfolio of assets that are being managed, or, by the consideration of critical business factors as measures for parameters that are relevant for better understanding the processes and realities of the financial scenarios.

\paragraph{}
Given the plethora of a large number of problem formulations, and a number of solutions for the portfolio optimization problem, identifying the exact formulation that would be applicable to the empirical data, and further taking a decision to select any one or a combination of a few, is a fairly complex decision that the portfolio manager would have to make on a continual basis. Furthermore, when the number of constraints to each formulation would vary, in such scenarios, identifying the set of constraints that are more suited to the nature of the empirical data and the behaviour of the financial assets; transforms the decision making process to even more challenging. Interestingly, it is often observed that when portfolio optimization problem formulations are not well-aligned to the empirical data, then, human bias and expertise does take control. With such real-world challenges, ensuring that the human bias relates to the numerical solutions in a more realistic way needs critical attention.

\paragraph{}
In light of the above challenges and complexities involved in the decision making process of financial portfolio optimization, this work focusses towards addressing the problem by combining a number of optimal solutions in a solution bank. It shall further investigate, analyse and optimize the solution with a bit more realistic global objective as directed by the individual investor or the financial institution such that the portfolio management process is realized using computationally guided assistance towards an informed decision making for financial investments.

\paragraph{}
This work suggests an agent based framework comprising of solver agents that contribute optimal solutions to a solution bank. Conceptually, these solver agents may represent various formulations of the problem, or mathematically oriented deterministic solvers, or information from published sources, or human knowledge encoded in numerical form, or a stochastic-search solver, or potentially, any entity that could provide a numerical solution to the weights that are associated to the assets in the portfolio. In principle, the solution bank shall then constitute of a variety of optimal solutions that meet various objective criteria under varying constraints, formulations and optimality reaching mechanisms.

\paragraph{}
Despite the presence of various mathematical formulations, the problem of selecting a specific solution instance across the entire spectrum of solutions has been a challenge in portfolio optimization. This work suggests the formulation of a global investment objective aligned according to the goals of the institution, and further optimize the allocation of weights by guiding the search for optimal solution by a super-agent solver, in the direction as guided by the set of optimal solutions in the solution bank. 

\paragraph{}
\begin{figure}[htb]
\centering
\includegraphics{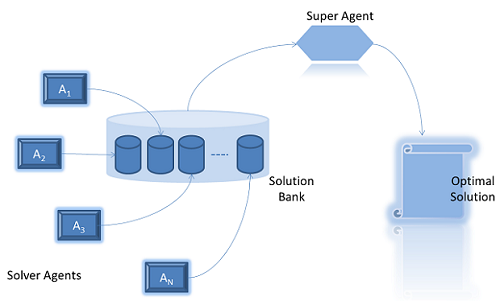}
\caption{Computationally Guided Agents Framework: Basic Model}
\label{fig:basic-framework}
\end{figure}

The notion of a super-agent solver is referred aesthetically and the super-ness is with regard to the enhanced capabilities of the search mechanism using the bank of optimal solutions. The super-agent solver may be a stochastic-search solver or a deterministic solver with multiple starting points. In principle, as long as the search is guided by a set of optimal solutions, the super-agent may be any of the solver agents as well, and as appropriate, it may be suitably interchanged with any solver agent.

\paragraph{}
This chapter shall further discuss the classes of solver agents, the solution bank orientation and computational methodology of reaching an optimal solution.

\section{Classes of Solver Agents}
\label{sec:solver-agents}

\paragraph{}
Entities that provide an optimal solution to any of the portfolio optimization formulation, by any of the solution strategies, or via knowledge from any internal/external sources of information are regarded as a potential agent that provides a solution, and is hence, further referred as a solver agent. Conceptually, based on the nature and formulation of the problem, and based upon the mechanism to obtain the solution, the solver agents are broadly classified amongst the following classes: human experience agents, domain knowledge agents, deterministic solver agents, and stochastic-search agents.

\paragraph{}
\begin{figure}[htb]
\centering
\includegraphics{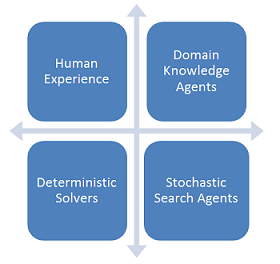}
\caption{Classes of Solver Agents}
\label{fig:solver-agents-classes}
\end{figure}

\paragraph{Human Experience Agents:}
Solver agents that constitute human experience from the financial markets, and those that represent numerical solutions encoded via human knowledge and sentiments are termed as the human experience agents. As well known, the process of portfolio optimization is a continual task that repeats over a period of time. This agent tends to include such periodic allocation information (from previous point in time, say previous month's allocation, or previous week's allocation etc.) as a potential solution in the solution bank; because, not making any changes while portfolio rebalancing, or, reverting back to a previous portfolio allocation, may well be considered as an optimal solution.

\paragraph{}
Human knowledge as encoded information from highly rated journals, trusted sources, syndicated services, or other published sources of information, when quantified, may also be considered as potential candidates for optimal solutions in the solution bank. Also, with the advent of the current day World Wide Web, and other electronic sources, human knowledge and experience are well-represented in social media information as user generated content in the form of blogs, user groups, discussion forums, mailing lists etc. Quantification of such information via Bayesian techniques, or other statistical methods, when compiled and integrated as numerical solutions, can be regarded as potential solutions.

\paragraph{Domain Knowledge Agents:}
Solver agents that represent knowledge from the financial domain by focusing on the domain-specific mathematical formulations are constituted as domain knowledge agents. These agents provide optimal solutions by focusing on varying investment objectives, and address business constraints from a more realistic point of view. Domain knowledge agents and domain expert agents shall interchangeably be referred, in this work, to symbolize the broad class of diverse mathematical formulations of the portfolio optimization problems of financial and statistical literature. 

\paragraph{}
As discussed in previous chapter, problem formulations that address the uncertainties in parameter estimation of the optimization problem via development of robust investment objective, or problem formulations that focus on strategic asset allocation mechanisms via risk-based asset allocation strategies focusing on risk-aligned investment objectives, and other mathematically driven formulations of the problem may be considered as domain knowledge agents. Deterministic (or stochastic) solutions of all such agents that have been formulated from financial literature are considered optimal towards their own formulation, but when viewed upon a broader perspective of investment management, each of them provide significant insights to the constraints on which these formulations are in place. A set of all such optimal solutions can potentially be taken for inclusion in the bank of solutions. 

\paragraph{Deterministic Solver Agents:}
Solver agents that focus on reaching the solution in a deterministic manner, based on methods and algorithms developed by mathematical rigour and proofs, are broadly labelled as deterministic solver agents. In principle, diverse formulations of the global investment objective may vary across linear programming, or quadratic programming, or even more general non-linear programming, or with the inclusion of integer constraints, it may be a mixed-integer non-linear programming problem, or may be solved using a sequential programming methodology etc. any many more such deterministic methods of mathematical programming. In general, solutions that are motivated across such disciplined methods are typically focused on the initial start of the numerical algorithm; and, may initiate from a single starting point or across multiple starting points. 

\paragraph{}
Approaches ranging across various single and multiple starting points, and algorithms that provide numerical approximations to the deterministic methods provide optimal solutions to the global investment objective formulation. The decision to consider multiple starting points may be regarded as a better one, but to decide the number of such points is often judged upon by heuristic insights. Larger number of starting points may span the entire search space and tends to be computationally expensive, whereas, a lower number of starting points may not be adequate enough. However, trying the multi-start approach with less number of starting points across multiple iterations with a cyclicity check is often considered useful. Optimal solutions obtained by the single-start and multi-start approaches of the deterministic methods are potential candidates for the solution bank, and can hence, be included. 

\paragraph{Stochastic-search Agents:}
Solver agents that focus on evolutionary computation techniques and swarm intelligence based methods for optimization problems are categorized as stochastic-search agents. These techniques emphasize on the use of random variables during the search process of an optimal solution. Computational methods that are inspired by biological mechanisms of evolution including reproduction, mutation, recombination and selection are considered as potential solver agents. These methods use iterative progression across a population of individuals, select them for guided random search and process them in parallel; referred as population-based meta-heuristic optimization methods. 

\paragraph{}
Other classes of algorithmic solution approach that believe in collective behaviour of systems which are decentralized and self-organized are referred as swarm intelligence based algorithms. Solutions obtained via numerical approximations from such nature-inspired algorithmic techniques are considered as potential candidates for the solution bank, and are hence included.

\paragraph{}
As studied during the literature survey, a number of computational methods have been tried to the portfolio optimization problem. The decision of selecting an algorithm is driven by the formulation and the author's know-how of the proposed methodology. The rationale for inclusion of solutions obtained via such computationally oriented stochastic-search methods is to gain deeper insights on the algorithm selection mechanism, especially when the problem formulations are complex and non-convex. The computationally guided mechanism of learning from a set of optimal solutions intends to strengthen the reasoning for the choice of using a particular algorithm towards a specific problem formulation

\section{Super-agent Solver}
\label{sec:super-agent}

\paragraph{}
The notion of a super-agent solver is an aesthetic conceptualization of any solver that would guide the search mechanism referring to the bank of optimal solutions. As mentioned earlier, it may be a mathematically oriented deterministic solver or a computationally oriented stochastic-search solver. This work does not discuss the notion of a deterministic solver and shall focus on the use of a stochastic-search mechanism as a potential super-agent solver. 

\paragraph{}
The field of Metallurgy presents one of the fundamental motivations for an algorithmic implementation of a stochastic-search technique by the annealing process in metals. It refers to a heat treatment to a solid state metal causing changes to its structural properties. It is followed by gradual cooling according to a specific schedule thereby obtaining a desired crystallized form. The metal is heated to a very high temperature to ensure free and random movement of the atoms of the solid state metal, and is further cooled, to ensure order amongst the atoms such that a thermal equilibrium is attained. By such an annealing process, the atoms position themselves corresponding to a desired crystallized form which happens only when the minimum energy configuration is reached; resulting in optimality of the solution. 

\paragraph{}
The technique mentioned above is referred as the Simulated Annealing methodology. It has the ability to stochastically disturb any current state of the solution and generate a neighbourhood state. The acceptance of the neighbourhood state as a potential solution is validated as per an acceptance criterion function. This process of disturbing the state, and accepting it further, is simulated across a number of Monte Carlo iterations and to avoid the algorithm to get stuck in the local optimum, the methodology of solution acceptance is guided by a temperature parameter as specified in the annealing schedule. Inclusion of such a temperature value, guides the algorithm towards the global optimum, and is hence, considered as a suitable candidate for a super-agent solver. 

\paragraph{}
\begin{algorithm}
\caption{Simulated Annealing Algorithm}
\begin{tabbing}
1. Start: Assign an initial state via some heuristic \\
2. Evaluate the objective function for initial state \\
3. \=Initialize: \\
\> a. Max. temperature value to a very high temperature \\
\> b. Max. no. of Monte Carlo iterations to a large number \\
4. \=While \{Max. temperature value $\geq$ frozen state temperature\} do \\
\>a. \=While \{no. of trials = 1 $\leq$ Max. no. of Monte Carlo iterations\} do \\
\>\>i. Generate a neighbourhood state with stochastic perturbation \\
\>\>ii. Evaluate the objective function for the neighbourhood state \\
\>\>iii. Apply the Metropolis rule and accept the generated neighbour \\
\>\>iv. Increment the no. of trials iterator \\
\>b. End of loop for Monte Carlo iterations/trials \\
\>c. Decrease the temperature according to the geometric annealing schedule \\
5. End of loop for temperature changes (or annealing schedules) \\
6. End of SA algorithm
\end{tabbing}
\end{algorithm}

Detailed discussion on the process of Simulated Annealing (SA) can be obtained from \citep{Kirkpatrick1983}. In essence, it is represented by the selection of the initial state, the neighbourhood generation mechanism, the acceptance criterion function, the annealing schedule and the stopping criterion. As a super-agent, this work considers a random initialization of the initial state, considers the Metropolis rule of the Metropolis-Hastings Algorithm as a function that accepts the generated solutions, follows a Geometric cooling schedule as a degenerating function for the annealing schedule, and a high temperature value with maximum number of Monte Carlo iterations for the termination condition of the SA algorithm. 

\paragraph{}
Conventionally, the generation of the neighbourhood state is performed by the generation of a uniform random number as a perturbation to the current state, and then the objective function is evaluated. This work proposes a methodology for generation of a neighbourhood state. The use of a solution bank tends to provide an increased guarantee for the solution as the search process is guided by the bank of optimal solutions and tends to reduce the randomness prevalent in the search mechanism.

\begin{figure}[htb]
\centering
\includegraphics{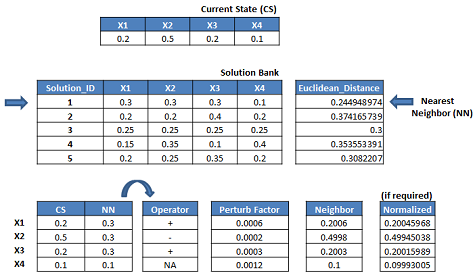}
\caption{Neighbourhood Generation: Solution Bank Approach}
\label{fig:solution-bank-approach}
\end{figure}

\paragraph{}
The approach, as shown in Figure ~\ref{fig:solution-bank-approach}, suggests calculation of a statistical distance measure between each solution instance in the solution bank with regard to the current state, and further selection of that solution instance which has minimum distance with that of the current state. Euclidean distance measure is used in this work. Then, the perturbation to the current state is based on the direction of further search guided by the orientation of the current state towards the nearest neighbour – across all solutions – from the bank.

\paragraph{}

\begin{figure}[htb]
\centering
\includegraphics{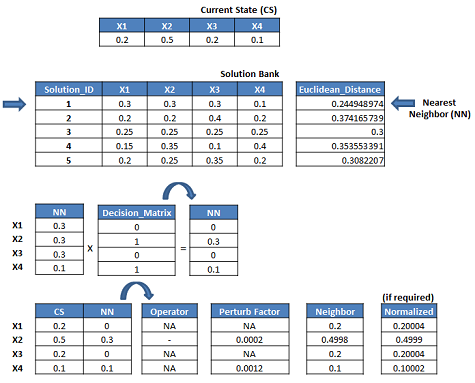}
\caption{Solution Bank Approach with Perturbation Strategy (Decision Matrix)}
\label{fig:solution-bank-approach-perturbation}
\end{figure}

While generating the neighbour, the way in which any current state is disturbed is responsible for the stochastic nature of the algorithm that facilitates in selection of the perturbed state as the possible neighbourhood state. This nature of stochastic perturbations to the parameters of the objective function for obtaining a feasible neighbour state is a critical aspect of SA technique. In traditional SA, stochastic perturbation to the current state is done to all decision variables of the objective function, whereas, this work further proposes the use of a perturbation strategy based on a decision matrix, as shown in Figure ~\ref{fig:solution-bank-approach-perturbation}, that shall control the perturbation mechanism of a select or all decision variables. The following perturbation strategies are proposed to maintain an optimal balance between the stochastic behaviours of various parameters:

\paragraph{Strategy I: Random Similar Perturbations:}
In this strategy, a decision vector of random zeros and ones is generated. It is based on a specific threshold that governs the uniform random number generator for filling the decision vector. The same decision vector is repeatedly applied to all parameters of the objective function. Application of similar perturbations to all parameters tends to reflect uniformity across parameters, and controls the nature of randomness while disturbing the current state. 

\paragraph{Strategy II: Random Varying Perturbations:}
In this strategy, the generation of a decision vector is replaced with the generation of a decision matrix where each parameter has a varying perturbation as compared to the other; however, the nature of generating the decision values of zeros and ones remains the same as the previous strategy. Application of varying perturbations to each parameter tends to reflect diversity across parameter perturbations and allows for free random movements from the current state. 

\paragraph{Strategy III: Perturb All:}
In this strategy, the use of any decision vector or matrix is avoided. It suggests perturbations to be applied to all parameters. This strategy is the simplest of all and reflects a direct perturbation based on the neighbourhood generation mechanism with the use of a solution bank.

\paragraph{}
In principle, the use of a decision matrix can be much more rigorous, and may be considered for careful stochastic perturbation. Possible alternative mechanisms for construction of the decision matrix include: (i) considering the distribution properties of the underlying assets in consideration, or (ii) generating a set of zeros and ones from a multivariate distribution, or (iii) enabling a suitable representation of the cardinality or holding constraints of the problem, if any, or (iv) provision for inclusion of the sectoral or risk-fraction constraints, if any, that would govern the fraction of risks that are associated with the assets, or (v) inclusion of heuristic information via deciding a threshold for the quantified information derived from the analysis of social media inputs for all assets in the portfolio.

\paragraph{}
This chapter has presented a computationally guided approach comprising of a variety of solver agents contributing their optimal solutions to a solution bank. It then proposed the construction of a problem formulation with a global investment objective. It further discussed an enhanced Simulated Annealing algorithm as a super-agent solver that represents a stochastic-search mechanism guided by the set of optimal solutions. It further proposed the use of a decision matrix for selective perturbation to the parameters of the objective function. The next chapter shall now present the detailed structure of the computational experiment and discuss the nature of the empirical data followed by an empirical validation of the proposed solution framework.

%% file: chapters/chap-4-empirical-validation-results.tex
\chapter{Empirical Validation and Results}
\label{ch:chap-4-empirical-validation-results}

\section{Experimental Setup}
\label{sec:experimental-setup}

\paragraph{}
The computationally guided agents approach as proposed in the previous chapter addresses many fundamental problems and concerns that were discussed in Chapters ~\ref{ch:chap-1-portfolio-optimization} and ~\ref{ch:chap-2-problem-formulations}. The proposed framework has been validated across various computational experiments that are discussed further in this chapter. The experimental setup fundamentally comprises of a number of solver agents, a conceptual solution bank, a global investment objective formulation and a super-agent solver that would computationally be guided by the solution bank whilst the stochastic-search process. This section shall provide detailed information on each of these components of the solution framework, and the next section shall be followed by the nature of the empirical data by preliminary exploratory analysis and problem scenarios.

\paragraph{Global Investment Objective:} 
The traditional Markowitz formulation with soft constraints on the returns, along with the budget constraint as mentioned in Section 2.1 is considered as the global objective. Since this work focusses on using a stochastic-search algorithm as a super-agent, the Markowitz formulation is modified. The return and budget constraints are addressed using the Penalty Functions approach. The penalty regularization parameters were experimented with varying values.

\paragraph{Markowitz Formulation with Penalty Functions}
\begin{equation}
\label{eqn: markopenalty}
\begin{aligned}
& \min & \sum^n_{i=1} \sum^n_{j=1} x_i x_j \, s_i s_j \, \rho_{ij} &&& \text{objective function} \\
& & + \alpha \, \sum^n_{i=1} [\max \{0, r_p - r_i x_i\}]^2 &&& + \beta \, (\sum^n_{i=1} x_i - 1)^2 &&& \\
& & 0 \leq x_i \leq 1, \; \forall i &&& \text{long-only constraint} \\
& \text{where, } & \rho_{ij} \in [-1, 1], \; \forall i, j &&& \text{correlation coefficient}
\end{aligned}
\end{equation}

\paragraph{Solver Agents:} 
The broad classes of solver agents represent human experience agents, domain knowledge agents, deterministic solver agents, and stochastic-search agents. Brief information about the characteristics of the solver agents is mentioned in Appendix ~\ref{ch:appendix-3}, and details are briefed below.

\paragraph{}
In the class of \textbf{human experience agents}, obtaining numerical solution for a select set of assets was not feasible and is therefore not considered in the experimental setup.

\begin{figure}[htb]
\centering
\includegraphics{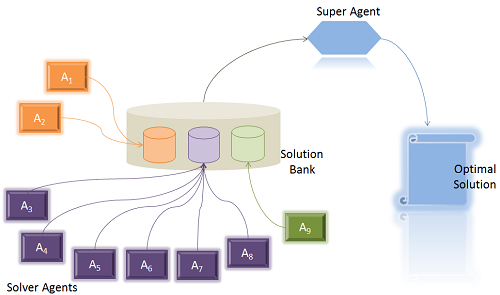}
\caption{Computationally Guided Agents Framework: Extended Model}
\label{fig:extended-framework}
\end{figure}

\paragraph{}
In the class of \textbf{domain knowledge agents}, the construction of a robust formulation of the problem addressing parameter uncertainty, and various problem formulations that address the risk-based asset allocation strategies as discussed in Chapter ~\ref{ch:chap-2-problem-formulations} were considered and deterministic solutions to these were obtained. These agents are represented in Purple (Solver Agents A3 to A8) in the extended model in Figure ~\ref{fig:extended-framework}. For the robust formulation, under the polyhedral uncertainty sets with the deterministic uncertainty model, \citep{Bertsimas2006} suggests the inclusion of an uncertainty constraint and an additional parameter to represent the budget of uncertainty. The uncertainty constraint controls the risk associated with the acceptable tolerance for constraint violation. This is reflected by the budget of uncertainty parameter in the objective function that refers to the deviation of such violations. 

\paragraph{}

The budget of uncertainty parameter is also related to the performance guarantee of the solution as it accounts for the constraint violations considered to achieve a suitable guarantee for the optimal solution. A higher performance guarantee requires a larger budget of uncertainty, whereas, a lower budget of uncertainty reflects a lesser guarantee for the obtained performance. Identifying a specific value for this parameter was not feasible; hence, solutions with varying performance guarantees were considered as potential candidates for the solution bank such that a suitable trade-off is maintained between the protection level of the constraint (by accounting for violations of constraints) and the degree of conservatism of the solution (by varying uncertainties in parameter estimates). Such variations in the budget of uncertainty parameter were accounted by obtaining fixed increments and were scaled across the nominal, conservative and robust strategies for parameter uncertainties.

\paragraph{}
In the class of \textbf{deterministic solver agents}, the global investment objective function was evaluated with both the single-start and multi-start approaches of the deterministic solver thereby solving the traditional Markowitz formulation as a Quadratic Programming problem. These agents are represented in Orange (Solver Agents A1 and A2) in the extended model in Figure ~\ref{fig:extended-framework}. The Markowitz formulation requires a risk-free rate of return (expected portfolio return) as input to the return constraint, and determining a suitable return value was not feasible. In this work, a number of expected portfolio return between the minimum and maximum return values (of the assets in the portfolio) were interpolated. Further, the deterministic solver agent was run a number of times corresponding to each interpolated expected portfolio return, in both the single-start and multi-start setup, so that portfolios representing various asset returns as portfolio returns could be considered as potential solutions for the bank.

\paragraph{}
In the class of \textbf{stochastic-search agents}, the Ant Colony Optimization (ACO) algorithm as adapted to continuous domains by \citep{Socha2008} and \citep{Liao2011} was considered. The ACO agent is represented in Green (Solver Agent A9) in the extended model in Figure ~\ref{fig:extended-framework}. In the stochastic setup, the global investment objective function of the Markowitz formulation was modified to incorporate the role of return and budget constraints using the Penalty functions approach. The modified formulation as presented in Equation ~\ref{eqn: markopenalty} was considered. Various values for the penalty regularization parameters were tried. Heuristic knowledge was also considered. Further, rather than selecting a single optimal solution for the bank of solutions, considering the stochastic nature of the solution methodology during the search process, various optimal weight structures that represent the optimal objective value were included in the solution bank. Fine-tuning of the parameters of ACO algorithm was carried out and a suitable combination is presented. 

\paragraph{}
As mentioned earlier, many other solver agents within each class of solvers could potentially be considered such that the optimal solutions are contributed to the solution bank. The underlying principle is to ensure that solutions that are optimal in any form, either via a problem formulation, or via a solution strategy, or via human expertise, or via numerical accuracy can all be considered as part of the solution bank such that the global investment objective is guided in a much informed manner. 

\paragraph{Solution Bank:} 
Considering the optimal solutions contributed by each solver agent as mentioned above and in Appendix ~\ref{ch:appendix-3}, the solution bank is a conceptual place holder for the set of optimal solutions. It comprises of the number of decision variables that are equivalent to the number of assets in the portfolio, and is hence, a matrix of optimal solutions where each row corresponds to an optimal solution and each column represents a decision variable as weights corresponding to each asset in the portfolio.

\paragraph{Super-Agent Solver:} 
As discussed in the previous chapter, the super-agent considered for the experimental setup is a stochastic-search methodology of Simulated Annealing (SA) algorithm with the solution bank approach. The inclusion of the decision matrix as a perturbation strategy is not considered as part of this empirical validation, and hence, all decision variables are perturbed.

\begin{figure}[htb]
\centering
\includegraphics{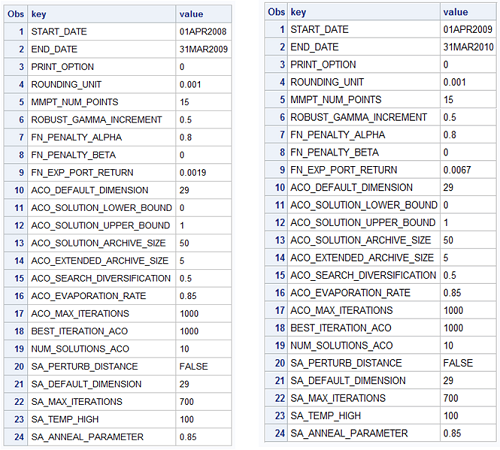}
\caption{Experimental Set-up: Solver Parameters}
\label{fig:solver-para}
\end{figure}

\begin{figure}[htb]
\centering
\includegraphics{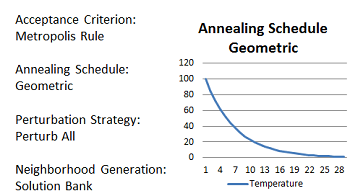}
\caption{Experimental Set-up: Super-Agent Configuration}
\label{fig:super-para}
\end{figure}

\paragraph{}
Figure ~\ref{fig:solver-para} contain the solver agent parameters and Figure ~\ref{fig:super-para} contains the super-agent configuration. These figures are placed at the end of this chapter. 

\newpage
\section{Nature of the Data}
\label{sec:nature-of-data}

\paragraph{}
The equities that constitute the BSE Sensex as on April, 2012 were chosen for the validation of the proposed solution framework. The daily close prices of the constituent assets were collected from the BSE Stock Exchange website. In the preliminary analysis phase, the behavior of the BSE Sensex was studied for the past 12 years. An increasing trend was observed for the daily close prices, except for the duration that represented the financial crisis of 2008.  

\paragraph{}
\begin{figure}[htb]
\centering
\includegraphics{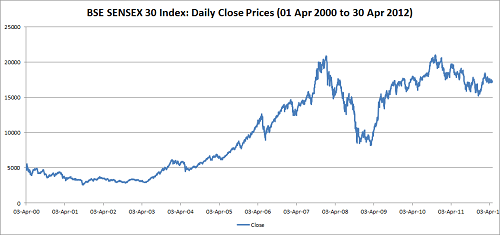}
\caption{BSE Sensex Index (2000-2012)}
\label{fig:bse-sensex-index}
\end{figure}

The downfall to recovery period from 2008 to 2010 as categorized by the downfall of the Sensex during the financial period 2008-2009 was observed, followed by an upward trend during the recovery of the Sensex in the financial period 2009-2010. The V-shaped trend of the market downward and upward trend was selected for validation of the proposed framework as representatives of the bearish and bullish market scenarios.

\paragraph{}
\begin{figure}[htb]
\centering
\includegraphics{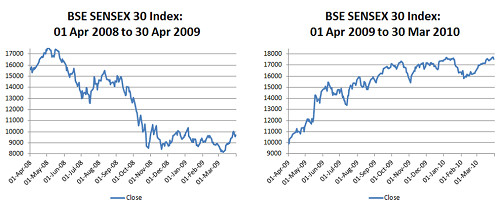}
\caption{Bearish \& Bullish Market Scenarios}
\label{fig:bearish-bullish}
\end{figure}

\paragraph{}
In general, BSE Sensex is composed of 30 equities; however, as per the selection of the above period, data sufficiency was obtained for 29 equities only. One of the equity i.e. Coal India Ltd. was observed to be listed on the stock exchange in later half of the year 2010, hence, daily close prices for the equity were not available. 

\paragraph{}
\begin{figure}[htb]
\centering
\includegraphics{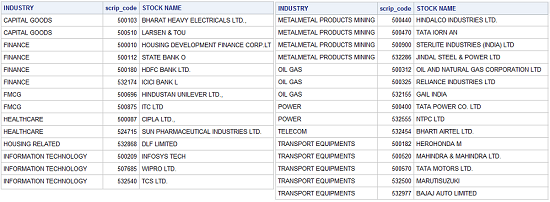}
\caption{Portfolio Constituents: 29 BSE Sensex Equities}
\label{fig:portfolio-constituents}
\end{figure}

For both the market scenarios, the arithmetic and log returns were computed on the daily and weekly close prices, but despite the lack of the symmetry property, the arithmetic returns on the daily close prices were considered realistic for the problem of portfolio optimization and was hence taken up as a suitable return measure. Also, the standard deviation on the daily and weekly arithmetic returns were computed, and the standard deviation on the daily arithmetic returns was considered as a suitable risk measure for the Markowitz formulation of the global investment objective function. 

\section{Scenario 1: Bearish Market}
\label{sec:bearish-market}

\paragraph{}
The BSE Sensex data corresponding to the financial year from 1st April, 2008 to 31st March, 2009 presents a declining trend representing the bearish market driven by the financial crisis of 2008. As discussed in the previous section, the daily close prices were considered for 29 equities. Return and risk measures, and the covariance matrix of asset returns were estimated for inputs to all the solver agents. 

\paragraph{}
Exploratory analysis, by means of a risk-return plot, of the risk-return measures presents that most of the assets in the portfolio provide a negative return. This behaviour is usually expected in the bearish market and is empirically validated as well. The correlation of the returns across all assets was studied, using the heat map visualization technique. It was observed that most of the assets provide a very high positive correlation. This affirms the belief of the bearish market in such a way that when the markets start falling, the investors prefer to sell all the equities as early as possible and thereby ensuring minimum loss due to further downfall, if any.

\paragraph{}
\begin{figure}[htb]
\centering
\includegraphics{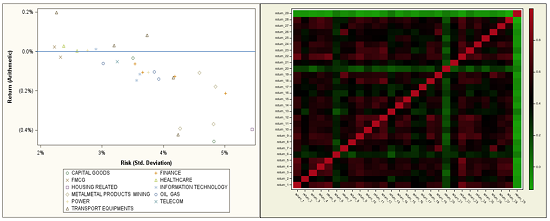}
\caption{Bearish Market Scenario: Exploratory Analysis}
\label{fig:bearish-explo}
\end{figure}

The expected behaviour of the empirical data during the bearish market scenario was validated by the exploratory analysis. The empirical data was further provided as input to the solver agents. Each of the solver agents were run, in series i.e. one after the other, and independent of each other. Optimal solutions from them were compiled in to the solution bank. The Simulated Annealing (SA) super-agent solver was executed to perform stochastic optimization by computational guidance from the solution bank. The search direction was hence guided by the optimal solutions. It was observed that the search process was inclined maximum number of times towards the optimal solution as obtained from the domain knowledge agent or robust formulation representing parameter uncertainty with the budget of uncertainty accountable as 0.5. And, upon convergence, SA inclined towards the same solution, thereby ensuring the belief of the local optimum to be the global optimum. 

\paragraph{}
\begin{figure}[htb]
\centering
\includegraphics{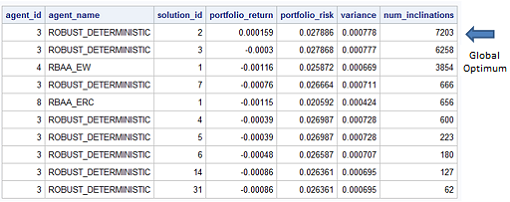}
\caption{Bearish Market Scenario: Solution Bank Inclinations}
\label{fig:bearish-incl}
\end{figure}

The SA algorithm further tried to improve upon the optimal solution from the solution bank towards a bit more finer optimality for a better risk-return combination as compared to the one provided by the Robust formulation. The weight structure seems to suggest that SA tried to allocate more weights in those industrial sectors that have performed better during the bearish market scenario. These industrial sectors include FMCG, Health care, and Transport equipments. It is often considered that certain industrial sectors form the backbone necessities of the fundamental infrastructure of the Country, and by such an increased allocation in those sectors, SA further emphasizes their importance, in particular, during the financial crisis. Also, it is observed that under the Markowitz formulation of diversification, SA tries to diversify suitably across industrial sectors and not across all assets in the portfolio, thereby providing an allocation that prefers a concentrated portfolio. 

\paragraph{}
\begin{figure}[htb]
\centering
\includegraphics{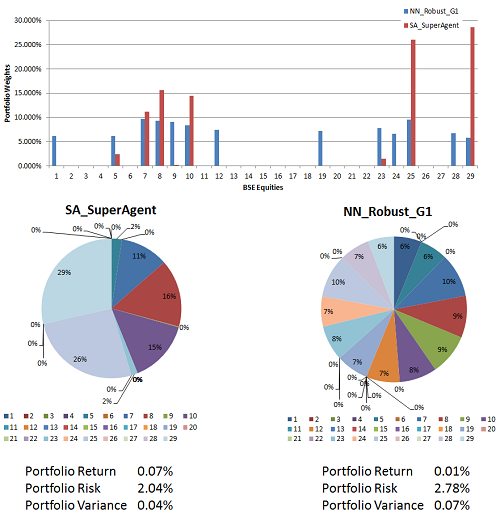}
\caption{Bearish Market Scenario: Experimental Results}
\label{fig:bearish-result}
\end{figure}

\section{Scenario 2: Bullish Market}
\label{sec:bullish-market}

The BSE Sensex data corresponding to the financial year from 1st April, 2009 to 31st March, 2010 presents an upward trend representing the bullish market driven by the recovery from the financial crisis of 2008. As discussed earlier, the daily close prices were considered for 29 equities. Return and risk measures, and the covariance matrix of asset returns were estimated for inputs to all the solver agents. 

\paragraph{}
Exploratory analysis, by means of a risk-return plot, of the risk-return measures presents that most of the assets in the portfolio provide a positive return. This behaviour is usually expected in the bullish market and is empirically validated as well. The correlation of the returns across all assets was studied, using the heat map visualization technique. It was observed that most of the assets provide a relatively lower positive correlation as compared to the empirical data corresponding to the bearish market. This affirms the belief of the bullish market in such a way that when the markets start rising, the investors prefer to take long positions in the equities but are equally cautious and sceptical about further rise, if at all any. 

\paragraph{}
\begin{figure}[htb]
\centering
\includegraphics{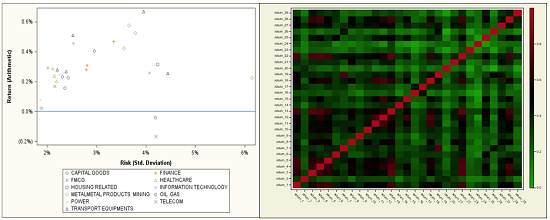}
\caption{Bullish Market Scenario: Exploratory Analysis}
\label{fig:bullish-explo}
\end{figure}

The expected behaviour of the empirical data during the bullish market scenario was validated by the exploratory analysis. The empirical data was further provided as input to the solver agents. Each of the solver agents were run, in series i.e. one after the other, and independent of each other. Optimal solutions from them were compiled in to another solution bank independent of the one compiled during the bearish scenario. The Simulated Annealing (SA) super-agent solver was executed to perform stochastic optimization by computational guidance from the solution bank. The search direction was hence guided by the optimal solutions. It was observed that the search process was inclined maximum number of times towards the optimal solution as obtained from the domain knowledge agent or robust formulation representing parameter uncertainty with the budget of uncertainty accountable as 1. And, upon convergence, SA inclined towards the robust formulation with the budget of uncertainty as 0.5. Such a behaviour thereby ensures the belief of the SA methodology with reference to not getting stuck in the local optimum and reaching out to the global optimum in the solution search space. 

\paragraph{}
\begin{figure}[htb]
\centering
\includegraphics{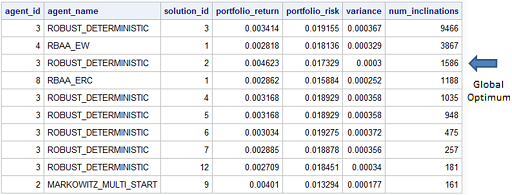}
\caption{Bullish Market Scenario: Solution Bank Inclinations}
\label{fig:bullish-incl}
\end{figure}

The SA algorithm further tried to improve upon the optimal solution from the solution bank towards a bit more finer optimality for a better risk-return combination as compared to the one provided by the Robust formulation. The weight structure seems to suggest that SA tried to allocate more weights in those industrial sectors that usually drive the economy in the bullish market scenario. These industrial sectors include Capital Goods, Finance, Metal, Metal products and Mining, and Transport equipments. By such an increased allocation in these sectors, SA further emphasizes their importance, in particular, for recovery from the financial crisis. Interestingly, it was again observed that under the Markowitz formulation of diversification, SA tried to diversify suitably across industrial sectors and not across all assets in the portfolio, thereby providing an allocation that preferred a concentrated portfolio over a well-diversified portfolio of all assets. 

\paragraph{}
\begin{figure}[htb]
\centering
\includegraphics{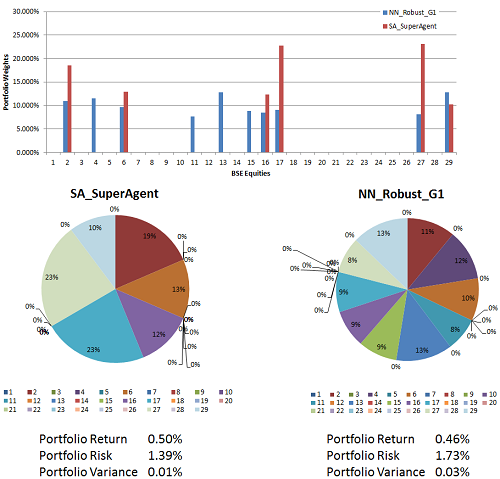}
\caption{Bullish Market Scenario: Experimental Results}
\label{fig:bullish-result}
\end{figure}

In this chapter, the proposed solution framework was validated across the bearish and bullish market scenarios. In both the circumstances, the Simulated Annealing methodology as adapted with the Solution Bank approach for neighbourhood generation was adequately validated thereby reaching to an optimal solution by suitable guidance from the a bank of optimal solutions and, more importantly, by keeping intact with the principles of stochastic-search optimization. The next chapter shall present the conclusions drawn from this work and discuss the future work and enhancements. 

\paragraph{}

\paragraph{}

%% file: chapters/chap-5-conclusions-future-work.tex
\chapter{Conclusion and Future Work}
\label{ch:chap-5-conclusions-future-work}

\paragraph{Conclusions}
\begin{itemize}
\item Portfolio optimization problem addresses a number of mathematical, statistical and financial formulations that are critical from various asset management perspectives. Identification of suitable business constraints considering the empirical data requires detailed understanding. This work proposed a computational model that allows addressing the above mentioned concerns by a guided knowledge-based search mechanism using a solution bank approach. The proposed framework was validated with the Bearish and Bullish market scenarios. 
\item Imposing various feasible business constraints to the portfolio optimization problems results in a very limited search space for search of optimal solution. Solving the formulation that best represents the global investment objective, and systematic experiments with each constraint or a group of constraints on the global objective would relatively search in a larger search space. A comparative assessment of the impact of the inclusion of such diverse constraints can be validated by the use of the proposed framework using solution bank approach.
\item The enhanced Simulated Annealing algorithm that leverage the computational guidance mechanism using solution bank based neighbourhood generation was considered as a potential super-agent solver, and was hence, empirically validated. In the bearish market scenario, it was observed that the local optimum results in global optimum, whereas, in the bullish market scenario, the global optimum was different from the local one. However, in both the scenarios, concentration of asset weights was observed towards certain critical industrial sectors.
\item Based on the solver agents that were considered, empirical investigations suggest that the empirical data is inclined towards Robust Formulation addressing parameter uncertainties. Hence, for the considered data, such a strategy towards portfolio optimization may be considered by the portfolio managers during a specified period. Such informed decisions are well-supported and guided by the novel technique as mentioned in this work.
\item To address human bias in asset management, or to address situations where problem formulations are not well aligned to empirical data, numerical solutions to the portfolio optimization formulations are related to human judgement via exploring the results that discuss the orientation of super-agent solver whilst the search. 
\end{itemize}

\newpage
\paragraph{Future Work}
\begin{itemize}
\item Problem formulations addressing risk-based and tactical asset allocation strategies, higher-order moments as parameters can be considered for finer approximations to empirical data.
\item Knowledge encoding mechanisms for constituent portfolio assets using advanced Bayesian techniques may be considered for potential candidates as human experience agents.
\item Harder realistic constraints may be imposed on problem formulations and may be considered as potential candidates for domain knowledge agents.
\item Other evolutionary computation techniques and swarm-intelligence methods may be algorithmically implemented as potential candidates for stochastic-search agents.
\item Various algorithmic techniques may be considered as super-agents that can leverage the benefits of the computationally guided mechanism of solution bank. Algorithmic modifications as done to Simulated Annealing algorithm for neighbourhood generation can be further studied.
\item Interchangeability of solver agent with that of the super-agent may be experimented.
\item The proposed framework may be empirically validated across neutral and stable market scenarios, and potentially, with various other asset classes (other than equities) that do not reflect higher variations and where parameter estimations are far more realistic.
\item The solution bank approach may be experimented with the out-of-sample data approach where simulated values for asset returns and risks are considered based on their distributional characteristics and various market behaviuor assumptions.
\item The random generation of the decision matrix can be guided by the distribution properties of the assets held in the portfolio, or can be used to represent the cardinality and/or sectoral constraints, or control the risk contributions of the assets, or can be estimated based on quantification of social media information for each asset in the portfolio.
\item The distance measure used to calculate the distance between the current state and all solutions in the solution bank may be explored beyond the Euclidean distance measure and may range across Manhattan distance, Minkowski distance, Chebyshev distance etc.
\item The significance of the correlations between the assets and the correlation behaviour of the solutions in the solution bank can be studied to provide enhanced computational guidance. 
\item The solution bank can be dynamically adapted and updated in parallel by the solver agents. 
\item The computational guidance mechanism may be experimented in an online cum dynamic mode where individual solver agents guide their optimal solutions via helping hands from other partnering solver agents by exploring the solution bank.

\end{itemize}

%% file: appendix/appendix-1.tex
\chapter{Return and Risk Measures}
\label{ch:appendix-1}

This appendix highlights a non-exhaustive list of the, absolute and relative, return and risk measures that are represented as parameters of the portfolio optimization problem.

\section{Return Measures}

\subsection{Absolute Return Measures}

\paragraph{Arithmetic Returns:}
It refers to the return on the asset as the ratio of the difference of the future price and current price with that of the current price.

\paragraph{Log Returns:}
It refers to the return on the asset as the logarithmic ratio of the future price and the current price of the asset.

\paragraph{Average Monthly Gain:}
It refers to arithmetic mean of the periods with a gain. It is calculated by summing the returns for gain periods, and dividing the total by the number of gain periods.

\paragraph{Average Monthly Loss:}
It refers to arithmetic mean of the periods with a loss. It is calculated by summing the returns for loss periods, and dividing the total by the number of loss periods.

\subsection{Relative Return Measures}

\paragraph{Up/Down Capture Ratio:}
The Up Capture Ratio is measured as an asset's compound return divided by the benchmark's compound return when the benchmark return increased. A higher value for this ratio is preferred. Whereas, the Down Capture Ratio is measured as an asset's compound return divided by the benchmark's compound return when the benchmark was down. A smaller value for the down ratio is preferred.

\paragraph{Up/Down Number Ratio:}
The Up Number Ratio is measured as the number of periods by which an asset's return increased, when the benchmark return increased, divided by the number of periods that the benchmark increased. A larger value for this ratio is preferred. Whereas, the Down Number Ratio is measured as the number of periods that an asset was down when the benchmark was down, divided by the number of periods that the benchmark was down. A smaller value for the ratio is preferred.

\paragraph{Up/Down Percentage Ratio:}
Also referred as the Proficiency Ratio. The Up Percentage Ratio is measured as the number of periods that an asset outperformed the benchmark when the benchmark increased, divided by the number of periods that the benchmark return increased. Whereas, the Down Percentage Ratio is measured as the number of periods that an asset outperformed the benchmark when the benchmark was down, divided by the number of periods that the benchmark was down. A larger value is preferred for both the Up and Down Percentage ratios.

\subsection{Absolute Risk-Adjusted Return Measures}

\paragraph{Sharpe Ratio:}
It is the measure of an asset's return relative to its risk. The return is defined as the asset's incremental average return over the risk-free rate. The risk is defined as the standard deviation of the asset's returns.

\paragraph{Calmar Ratio:}
It is the return to risk ratio. The return is defined as the compound annualized return over the previous years, and the risk is defined as the maximum draw down (in absolute terms) over the previous years.

\paragraph{Sterling Ratio:}
It is the return to risk ratio. The return is defined as the compound annualized return over the previous years, and the risk is defined as the average yearly maximum draw down over the previous years subtracted by an arbitrary 10\%.

\paragraph{Sortino Ratio:}
It is the return to risk ratio. The return is defined as the incremental compound average period return over a minimum acceptable return (MAR), and the risk is defined as the downside deviation below the MAR. It is a modification of the Sharpe ratio but penalizes only those returns that fall below a specified target, or required rate of return, while the Sharpe ratio penalizes both upside and downside volatility equally.

\subsection{Relative Risk-Adjusted Return Measures}

\paragraph{Annualized Alpha:}
It measures an asset's value added relative to any benchmark. It is referred as the intercept of the regression line.

\paragraph{Jensen Alpha:}
It measures the extent to which an asset has added value relative to a benchmark. It is equal to an asset's average return in excess of the risk-free rate of return, subtracted by beta times the benchmark's average return in excess of the risk-free rate.

\paragraph{Treynor Ratio:}
It is similar to the Sharpe ratio, but it refers to beta as the volatility measure rather than standard deviation. The return is defined as the incremental average return of an asset over the risk-free rate of return. The risk is defined as an asset's beta relative to a benchmark. A larger value is preferred.

\paragraph{Information Ratio:}
It measures an asset's active premium divided by the asset's tracking error. It relates the degree to which an asset beats the benchmark to the consistency by which the asset has beaten the benchmark.

\section{Risk Measures}

\subsection{Absolute Risk Measures}

\paragraph{}
Standard Deviation over weekly, monthly and yearly asset returns, Standard Deviation for Gain and Loss periods, Gain-Loss Ratio (that measures an asset's average gain in a gain period divided by the asset's average loss in a loss period), Skewness, Kurtosis etc. and other higher moments are considered as Absolute Risk Measures.

\subsection{Relative Risk Measures}

\paragraph{Beta:}
It represents the slope of the regression line. It measures an asset's risk relative to the market as a whole (where, the 'market' is any index or investment). Beta describes the sensitivity of the asset to broad market movements. 

\subsection{Tail Risk Measures}

\paragraph{Value at Risk (VaR):}
It is the maximum loss that can be expected within a specified holding period and a specified confidence level. It assumes that the assets are normally distributed. In order to relax the normality assumption, the Modified Value at Risk is calculated in the same manner as VaR. The modified VaR 'corrects' the VaR using the calculated skewness and kurtosis of the distribution of returns.

\paragraph{Expected Tail Loss (ETL):}
Also referred as the Conditional Value at Risk (CVaR). It is the average expected loss beyond VaR; and, is used for risk budgeting. It can be interpreted as the expected shortfall assuming VaR as the benchmark. It is a downside risk measure that can recognize diversification opportunities. It is sub-additive and is used to aggregate/decompose risk at various portfolio levels.

\paragraph{STARR Ratio:}
Abbreviated as Stable Tail Adjusted Return Ratio. The evaluation of risk adjusted performance is an alternative to the Sharpe Ratio, but STARR takes into account the drawback of standard deviation as a risk measure, which penalizes not only for upside but for downside potential as well. It employs the Expected Tail Loss of the asset returns for the performance adjustment.

\paragraph{Rachev Ratio:}
It is a reward-to-risk measure and defined as the ratio between the expected tail loss of the opposite of the excess return at a given confidence level and the expected tail loss of the excess return at another confidence level. It is the expected tail return (ETR) divided by expected tail loss (ETL). Where, ETL is the average of the returns that exceed the VaR number in the left tail, and ETR is the average of say 5\% of returns in the right tail at say 95\% confidence level.

%% file: appendix/appendix-2.tex
\chapter{Business \& Technical Constraints}
\label{ch:appendix-2}

This appendix highlights a non-exhaustive list of business constraints that are crucial for financial portfolio optimization. In practice, many more constraints are applicable. 

\paragraph{Budget Constraint:}
It represents the combinations of assets that can be held in the portfolio with the amount of money available for investment. It enforces that the total proportion of weights to be allocated should always equal 100\% of the total investment.

\paragraph{Return Constraint:}
It represents the expected return from the portfolio. It enforces that the return expected from the portfolio equals the expected return from any risk-free asset. Meeting this criteria using deterministic solvers might lead to higher number of iterations. To overcome convergence issues, it is often transformed to a soft constraint.

\paragraph{Long-only Constraint:}
It allows for taking long positions only while portfolio rebalancing. It rules out the possibility of short positions; i.e. assets shall not be sold while managing the portfolio. Mathematically, it disallows for negative weights.

\paragraph{Turnover Constraint:}
Also referred as Purchase and/or Sale Constraint. It represents the amount of money (including purchase and sales of assets) that can be traded. It imposes upper bounds on the variation of the asset holdings from one period to the next.

\paragraph{Holding Constraint:}
Also referred as Cardinality Constraint. It limits the number of assets that can be held in the portfolio. It allows for reduced portfolio re-weighting costs and efficient monitoring by dealing with a limited number of assets. 

\paragraph{Trading Constraint:}
It limits the number of assets in the portfolio that can be traded while managing the portfolio. It is usually enforced to reduce the burden of transaction costs, trading costs, and brokerages. 

\paragraph{Risk Factor Constraint:}
Also referred as Risk Fraction Constraint. It represents the fraction of portfolio risk that can be attributed to each asset. It is useful for creating risk parity portfolios and efficient management of risk allocations in the portfolio.

\paragraph{Tracking Error Constraint:}
Also referred as Benchmark Exposure Constraint. It represents the amount of deviation a portfolio can have when compared to any standardized benchmark. It is often imposed by portfolio managers to compare the performance of their portfolio.

\paragraph{Round Lot Constraint:}
Financial trading is mostly done by integer holdings of assets. When asset prices are given for lots, portfolio balancing is done accordingly. Positions that have non-integer values are excluded from the scope of this constraint.

\paragraph{Volatility Constraint:}
It represents the expected volatility of the portfolio. It enforces that the portfolio volatility lie between certain bound constraints. It constrains the riskiness of the portfolio. In practice, multiple portfolio volatility checks are enforced.

\paragraph{Distance Constraint:}
It limits the distance between the portfolio and some target portfolio i.e. the amount of money required to trade from one portfolio to the other. In principle, portfolio distance can be computed in many ways. The sum of absolute differences in weights is usually practised.

\paragraph{Closing Position Constraint:}
A position is considered closed if an asset is in an existing portfolio but not in the revised portfolio. This constraints controls the number of positions (incl. long and short) that are closed.

\paragraph{Cost Constraint:}
If transaction costs are applicable, then bounds on such costs are represented by the cost constraint.

\paragraph{Largest Weight Constraint:}
It limits the sum of a fixed number of the largest weights, irrespective of the assets that have them. It is often confused with the constraint that controls the bounds on the asset holdings. 

\paragraph{Sectoral Constraint:}
It represents the control on the amount of investment in a particular industrial sector. Broadly, all constraints that can be enforced on individual assets can be applied to industrial sectors.

\paragraph{Forced Trade Constraint:}
It enforces trades to be made of certain select assets with a specified size. It is different from threshold constraint that limits only the trade size.

%% file: appendix/appendix-3.tex
\chapter{Characteristics of Solver Agents}
\label{ch:appendix-3}

\paragraph{}
\begin{tabular}{ | l | p{10cm} |}
\hline
Solver Agent ID 		& A1 \\ \hline
Solver Agent Type 		& Deterministic Solver Agent \\ \hline
Solver Agent Name 		& MARKOWITZ\_SINGLE\_START \\ \hline
Solver Orientation 		& Deterministic Optimization \\ \hline
Solver Method 			& QP Solver (Quadratic Programming)	\\ \hline
Solver Details 			& Infeasible Primal-Dual Predictor-Corrector Interior Point Algorithm \\ \hline
Solver References 		& Chapter 8: The Quadratic Programming Solver from SAS/OR 9.3 User's Guide: Mathematical Programming \\ \hline
No. of Solutions 		& 15 \\ \hline
Dependent Parameters 	& MMPT\_NUM\_POINTS \\ \hline
Remarks					& The number of expected portfolio return values that were interpolated between the minimum and maximum return on any asset in the portfolio were set to 15 using the above parameter. \\ \hline
\end{tabular}

\paragraph{}
\begin{tabular}{ | l | p{10cm} |}
\hline
Solver Agent ID 		& A2 \\ \hline
Solver Agent Type 		& Deterministic Solver Agent \\ \hline
Solver Agent Name 		& MARKOWITZ\_MULTI\_START \\ \hline
Solver Orientation 		& Deterministic Optimization \\ \hline
Solver Method 			& NLP Solver with MS (Non-Linear Programming Solver with Multi-start) \\ \hline
Solver Details 			& Infeasible Primal-Dual Interior Point Algorithm \\ \hline
Solver References 		& Chapter 7: Nonlinear Programming Solver from SAS/OR 9.3 User's Guide: Mathematical Programming \\ \hline
No. of Solutions 		& 15 \\ \hline
Dependent Parameters 	& MMPT\_NUM\_POINTS \\ \hline
Remarks					& The number of expected portfolio return values that were interpolated between the minimum and maximum return on any asset in the portfolio were set to 15 using the above parameter. \\ \hline
\end{tabular}

\paragraph{}
\begin{tabular}{ | l | p{10cm} |}
\hline
Solver Agent ID 		& A3 \\ \hline
Solver Agent Type 		& Domain Knowledge Agent / Domain Expert Agent \\ \hline
Solver Agent Name 		& ROBUST\_DETERMINISTIC \\ \hline
Solver Orientation 		& Deterministic Optimization \\ \hline
Solver Method 			& OPTLP Solver (Linear Programming) \\ \hline
Solver Details 			& Dual Simplex Solver (Two Phase Simplex Algorithm) \\ \hline
Solver References 		& Chapter 10: The OPTLP Procedure from SAS/OR 9.3 User's Guide: Mathematical Programming \\ \hline
No. of Solutions 		& 59 (0 to 29 with an increment of 0.5) \\ \hline
Dependent Parameters 	& ROBUST\_GAMMA\_INCREMENT \\ \hline
Remarks					& The budget of uncertainty parameter is scaled from 0 to the number of assets constituent in the portfolio with a fixed increment of 0.5 \\ \hline
\end{tabular}

\paragraph{}
\begin{tabular}{ | l | p{10cm} |}
\hline
Solver Agent ID 		& A4 \\ \hline
Solver Agent Type 		& Domain Knowledge Agent / Domain Expert Agent \\ \hline
Solver Agent Name 		& RBAA\_EW (Risk-based Asset Allocation: Equally Weighted Portfolio) \\ \hline
Solver Orientation 		& None \\ \hline
Solver Method 			& None \\ \hline
Solver Details 			& None \\ \hline
Solver References 		& None \\ \hline
No. of Solutions 		& 1 \\ \hline
Dependent Parameters 	& Number of Equities \\ \hline
Remarks					& Equal-weights are assigned based on the number of equities. No numerical optimization is involved. \\ \hline
\end{tabular}

\paragraph{}
\begin{tabular}{ | l | p{10cm} |}
\hline
Solver Agent ID 		& A5 \\ \hline
Solver Agent Type 		& Domain Knowledge Agent / Domain Expert Agent \\ \hline
Solver Agent Name 		& RBAA\_GMV (Risk-based Asset Allocation: Global Minimum Variance) \\ \hline
Solver Orientation 		& Deterministic Optimization \\ \hline
Solver Method 			& QP Solver (Quadratic Programming) \\ \hline
Solver Details 			& Infeasible Primal-Dual Predictor-Corrector Interior Point Algorithm \\ \hline
Solver References 		& Chapter 8: The Quadratic Programming Solver from SAS/OR 9.3 User's Guide: Mathematical Programming \\ \hline
No. of Solutions 		& 1 \\ \hline
Dependent Parameters 	& None \\ \hline
Remarks					& None \\ \hline
\end{tabular}

\paragraph{}
\begin{tabular}{ | l | p{10cm} |}
\hline
Solver Agent ID 		& A6 \\ \hline
Solver Agent Type 		& Domain Knowledge Agent / Domain Expert Agent \\ \hline
Solver Agent Name 		& RBAA\_MDP\_DR (Risk-based Asset Allocation: Most-Diversified Portfolio using Diversification Ratio) \\ \hline
Solver Orientation 		& Deterministic Optimization \\ \hline
Solver Method 			& NLP Solver (Non-Linear Programming) \\ \hline
Solver Details 			& Infeasible Primal-Dual Interior Point Algorithm \\ \hline
Solver References 		& Chapter 7: Nonlinear Programming Solver from SAS/OR 9.3 User's Guide: Mathematical Programming \\ \hline
No. of Solutions 		& 1 \\ \hline
Dependent Parameters 	& None \\ \hline
Remarks					& None \\ \hline
\end{tabular}

\paragraph{}
\begin{tabular}{ | l | p{10cm} |}
\hline
Solver Agent ID 		& A7 \\ \hline
Solver Agent Type 		& Domain Knowledge Agent / Domain Expert Agent \\ \hline
Solver Agent Name 		& RBAA\_MDP\_RW (Risk-based Asset Allocation: Most-Diversified Portfolio using Risk-Weighted constraint) \\ \hline
Solver Orientation 		& Deterministic Optimization \\ \hline
Solver Method 			& NLP Solver (Non-Linear Programming) \\ \hline
Solver Details 			& Infeasible Primal-Dual Interior Point Algorithm \\ \hline
Solver References 		& Chapter 7: Nonlinear Programming Solver from SAS/OR 9.3 User's Guide: Mathematical Programming \\ \hline
No. of Solutions 		& 1 \\ \hline
Dependent Parameters 	& None \\ \hline
Remarks					& None \\ \hline
\end{tabular}

\paragraph{}
\begin{tabular}{ | l | p{10cm} |}
\hline
Solver Agent ID 		& A8 \\ \hline
Solver Agent Type 		& Domain Knowledge Agent / Domain Expert Agent \\ \hline
Solver Agent Name 		& RBAA\_ERC (Risk-based Asset Allocation: Equally-weighted Risk Contribution Strategy) \\ \hline
Solver Orientation 		& Deterministic Optimization \\ \hline
Solver Method 			& SQP Solver (Sequential Programming) \\ \hline
Solver Details 			& Iterative procedure using Augmented Lagrangian Penalty Functions as a Merit function solves QP sub-problem to obtain a search direction, followed by a Line-Search Algorithm along resulting search direction. \\ \hline
Solver References 		& Chapter 13: The Sequential Programming Solver from SAS/OR 9.2 User's Guide: Mathematical Programming
 \\ \hline
No. of Solutions 		& 1 \\ \hline
Dependent Parameters 	& None \\ \hline
Remarks					& None \\ \hline
\end{tabular}

\paragraph{}
\begin{tabular}{ | l | p{10cm} |}
\hline
Solver Agent ID 		& A9 \\ \hline
Solver Agent Type 		& Stochastic-search Agent \\ \hline
Solver Agent Name 		& MARKOWITZ\_PENALTY \\ \hline
Solver Orientation 		& Stochastic Optimization \\ \hline
Solver Method 			& ACO-R Algorithm \\ \hline
Solver Details 			& Ant Colony Optimization (ACO) adapted to Continuous Domains (R) \\ \hline
Solver References 		& \citep{Socha2008} and \citep{Liao2011} \\ \hline
No. of Solutions 		& 10 \\ \hline
Dependent Parameters 	& BEST\_ITERATION\_ACO, NUM\_SOLUTIONS\_ACO \\ \hline
Remarks					& BEST\_ITERATION\_ACO represents the best solution index across ACO iterations, and NUM\_SOLUTIONS\_ACO represents the number of best solutions to be considered for the solution bank.  \\ \hline
\end{tabular}